\documentclass[journal]{IEEEtran}
\usepackage[perpage,symbol]{footmisc}
\usepackage{cite}
\usepackage[dvipsnames]{xcolor}
\usepackage{rotating}
\usepackage {amsmath}
\usepackage {amssymb}
\usepackage{algorithm,algpseudocode}
\usepackage {graphicx}
\usepackage [font=small]{caption}
\usepackage {subcaption}
\usepackage {url}
\usepackage{multirow}
\usepackage{epstopdf}
\usepackage{diffcoeff}
\usepackage{mathtools}
\usepackage{mathptm}
\usepackage{fixltx2e}
\usepackage[T1]{fontenc}
\graphicspath {{../Figures}{../Graphs}}
\usepackage{breqn}
\usepackage{gensymb}
\usepackage[english]{babel}
\usepackage{microtype}
\usepackage{booktabs}

\DeclarePairedDelimiter\ceil{\lceil}{\rceil}

\makeatletter
\let\@copyrightspace\relax
\makeatother

\def\BibTeX{{\rm B\kern-.05em{\sc i\kern-.025em b}\kern-.08em
    T\kern-.1667em\lower.7ex\hbox{E}\kern-.125emX}}

\begin{document}
\title{Designing Efficient and High-performance AI Accelerators with Customized STT-MRAM}
\author{Kaniz~Mishty,~
        and~Mehdi~Sadi,~\IEEEmembership{Member~,~IEEE}
\thanks{This work was supported in part by the IGP grant from Auburn University, Auburn, AL.}
\thanks{Kaniz Mishty and Mehdi Sadi are with the Department of Electrical and Computer Engineering, Auburn University, Auburn, AL 36849, USA (e-mail: kzm0114@auburn.edu, mehdi.sadi@auburn.edu)}
\thanks{Manuscript received April 05, 2021; revised XX XX, 2021; accepted XX XX, 2021.}}
\markboth{IEEE Transactions on Very Large Scale Integration (VLSI) Systems,~Vol.~XX, No.~X, August~202X}%
{Sadi \MakeLowercase{\textit{et al.}}: STT-AI: Designing Efficient and high-performance AI Accelerators with Customized STT-MRAM}
\maketitle

\begin{abstract}
In this paper, we demonstrate the design of efficient and high-performance AI/Deep Learning accelerators with customized STT-MRAM and a reconfigurable core. Based on model-driven detailed design space exploration, we present the design methodology of an innovative scratchpad-assisted on-chip STT-MRAM based buffer system for high-performance accelerators. Using analytically derived expression of memory occupancy time of AI model weights and activation maps, the volatility of STT-MRAM is adjusted with process and temperature variation aware scaling of thermal stability factor to optimize the retention time, energy, read/write latency, and area of STT-MRAM. From the analysis of modern AI workloads and accelerator implementation in 14nm technology, we verify the efficacy of our designed AI accelerator with STT-MRAM (\textit{STT-AI}). Compared to an SRAM-based implementation, the STT-AI accelerator achieves 75\% area and 3\% power savings at iso-accuracy. Furthermore, with a relaxed bit error rate and negligible AI accuracy trade-off, the designed \textit{STT-AI Ultra} accelerator achieves 75.4\%, and 3.5\% savings in area and power, respectively over regular SRAM-based accelerators.

\end{abstract}

				
\section{Introduction}
\label{sec:Intro}
The demand for Deep Learning and Artificial Intelligence (AI) is growing at a rapid pace across a wide range of applications such, as self-driving vehicles, image and voice recognition, medical imaging and diagnosis, finance and banking,  defense operations, etc. Because of these data-driven analytics and AI boom, demands in deep learning and AI will emerge at both data centers and the edge \cite{market,aa1,tpu}.  In a recent market research \cite{market}, it has been reported that AI-related semiconductors will see a growth of about 18 percent annually over the next few years - five times greater than the rate for non-AI applications. By 2025, AI-related semiconductors could account for almost 20 percent of all semiconductor demand, which would translate into about \$67 billion in revenue \cite{market}. As a result, significant R\&D efforts in developing AI accelerators - optimized to achieve much higher throughput in deep learning compared to GPUs  -  are underway from academia, big techs, as well as startups \cite{aa1}. In AI technology innovation and leadership, high-throughput AI accelerator hardware chips will serve as the differentiator \cite{tpu,market}.

On-chip memory capacity plays a significant role in the performance and energy efficiency of AI tasks \cite{aa1,eyeriss,DNPU,tpu}. In AI accelerator, off-chip Dynamic Random-Access Memory (DRAM) accesses can take 200 times and 10 times more energy compared to the local register file and global buffer memory, respectively \cite{eyeriss}.  Larger on-chip buffer memory is needed to minimize DRAM accesses, and it can improve the energy efficiency and speed of the accelerator. However, conventional Static Random-Access Memory (SRAM) based solutions suffer from area constraints and leakage power at advanced technology nodes \cite{imec, memory_trend}, which is a major concern for the energy-constraint IoT domain. STT-MRAM has the potential to replace SRAM as the global buffer in high-performance AI accelerators that require large on-chip memory \cite{cb,tpu}. For AI accelerators used in inference-only applications, the pre-trained weights need to be stored on-chip. As conventional embedded Flash storage suffers from scalability and reliability issues at advanced nodes \cite{ memory_trend}, emerging memory-based solutions are required for AI accelerators. As analyzed in detail in \cite{ser}, because of weight reuse in Deep Learning, radiation-induced soft errors in the memory block of the accelerator can impact the accuracy of AI models. This is especially a concern for safety-critical applications such as autonomous vehicles with rigid FIT requirements \cite{ser}, and STT-MRAM can be a better option for these types of applications.

At scaled technologies (e.g., 10nm and newer), static energy loss from the high leakage current dominates the overall energy dissipation in DRAM and SRAM technologies \cite{memory_trend}. Although Trench cap based embedded DRAM (eDRAM) has a higher density compared to SRAM, the leakage power and scaling challenges of eDRAM at advanced process nodes make it less competitive in the future technology roadmap \cite{memory_trend}. Beyond 28nm node eFlash faces scaling challenges, and eMRAM technology becomes superior over eFlash because of its lower write voltage and energy, higher endurance, lower area, and faster read/write time \cite{samsung}. The emerging resistive RAM (RRAM) and Phase Change (PCM) based cross-point memory suffers from endurance, reliability and variability problems \cite{dac_19, memory_trend}. Among all the emerging embedded memory technologies, STT-MRAM is one of the most  promising due to its high energy efficiency, write endurance (e.g., more than 1 million cycles), high cell density, high-temperature data retention capability, operating voltage comparable with CMOS logic, and immunity to soft errors \cite{memory_trend,imec,stt_ram_tsmc1,intel,stt_ram_tsmc2,samsung_19,ibm_2019}. Moreover, STT-MRAM is higly compatible with CMOS and requires only 2 to 6 extra masks in the backend-of-the-line (BEOL) process \cite{intel,imec}. Because of the leakage power issue, beyond a certain memory size, embedded MRAM becomes more energy efficient compared to SRAM \cite{imec}.

While performing Deep Learning/AI tasks, the throughput of the AI accelerator primarily depends on, (i) the number of Processing Elements (PE), and (ii) the size of on-chip buffer memory \cite{aa1,tpu}. As a result, area-efficiency is of paramount importance for AI accelerators, and the critical design goal is to increase PE density and on-chip memory capacity. Because of compact size ($~6F^2$ of STT-MRAM vs. $~100F^2$ of SRAM \cite{destiny,stt_model}), STT-MRAM has the potential to outperform conventional SRAM as the on-chip memory in accelerators. At iso-memory capacities, the MRAM module occupies much lower area compared to SRAM \cite{imec}. Additionally, for power constraint mobile/edge/IoT applications, STT-MRAM based AI accelerators can significantly minimize static power compared to SRAMs. However, the higher write energy and write latency of conventional eMRAM can be a deterrent in their full adoption in AI accelerators. In this paper, we present a methodology to design efficient AI accelerators with customized STT-MRAM that can provide high bit cell density while still ensuring fast write speed and decreased write energy. We achieve this feat by analyzing the volatility requirement of weight and input/output feature-map (ifmap/ofmap) data on-chip, and scaling the eMRAM's retention time accordingly without incurring unacceptable bit error rates.

To the best of our knowledge, this is the first extensive work on designing AI/Deep Learning accelerators with STT-MRAM based on-chip memory systems. The key contributions and highlights of this paper are,

\begin{itemize} 

\item We present an innovative runtime reconfigurable core design that can be optimized for both dot products of convolution layers and matrix multiplications of fully connected layers.

\item  We derive the analytical expressions of occupancy times of weights and input/output feature maps in the global memory of the AI accelerator between different stages (i.e., Conv. layer followed by Conv., Conv. layer followed by Fully-Connected (FC) layer, and FC-FC) of AI/Deep Learning operation. Guided by this data activity duration, we scale the retention time of STT-MRAM and customize the design for application as the global buffer memory in energy-efficient AI accelerator. We consider Process variations and runtime Temperature  fluctuations in this scaling procedure to ensure negligible read/write Bit Error Rates (BER) and retention failures across all corners.

\item Based on detailed design space exploration using state-of-the-art AI/Deep Learning models, an AI accelerator system and MRAM technology co-design framework is presented with the key innovations, - (i) Optimizes STT global buffer size to minimize DRAM accesses. (ii) A novel scratchpad-assisted STT-MRAM based global buffer architecture is presented to minimize the writes to the MRAM by bypassing writes of the partial ofmaps to the scratchpad.  (iii) For inference-only tasks, to store the trained weights a specially customized embedded STT-MRAM - as a Flash replacement - with optimized retention time (e.g., 3 to 4 years) and robust BER is used.

\item To further improve the energy and area efficiency, we exploited the inherent error tolerance of Deep Learning/AI models and created two STT-MRAM banks for the global buffer. For the first bank, the thermal stability factor is scaled further to a relaxed BER, and the less critical half of the weights/fmap bits (e.g., LSB groups) are assigned to this memory block. The second bank has scaled retention time with a robust BER, and the other remaining half of the bits (e.g., MSB groups) are stored in this bank. 
\end{itemize}

The rest of the paper is organized as follows. The background is discussed in Section II. STT-MARM based optimum AI/Deep learning accelerator design methodology is presented in Section III.  The AI accelerator-aware eMRAM technology co-design methodology is presented in Section IV. We present Simulation Results in Section V, Related work in Section VI, and Conclusions in Section VII.

\section{Background}
\label{sec: Background}
\subsection{Deep Neural Networks}
At the core of Deep Learning/AI is the Deep Neural Network (DNN). Modern state-of-the-art DNN consists of stacks of Convolution layers to extract the objects' features and a few Fully Connected layers at the end to classify them. Convolutions are element-wise dot products between matrix (or vector) and matrix. In convolution, kernels convolve over input feature maps (ifmap) to extract embedded features and generate the output feature maps (ofmap) by accumulating the partial sums (psums) as shown in Fig. \ref{fig:conv_op}. Each fmap and filter is a 3D structure consisting of multiple 2D planes, and a batch of 3D fmaps is processed by a group of 3D filters in a layer. Activation functions (e.g., ReLU) operate on the results before they go to the MaxPooling layer. The computations of a convolutional layer can be expressed as:
\begin{figure}[h]
	\centering	
	\includegraphics[scale=.89]{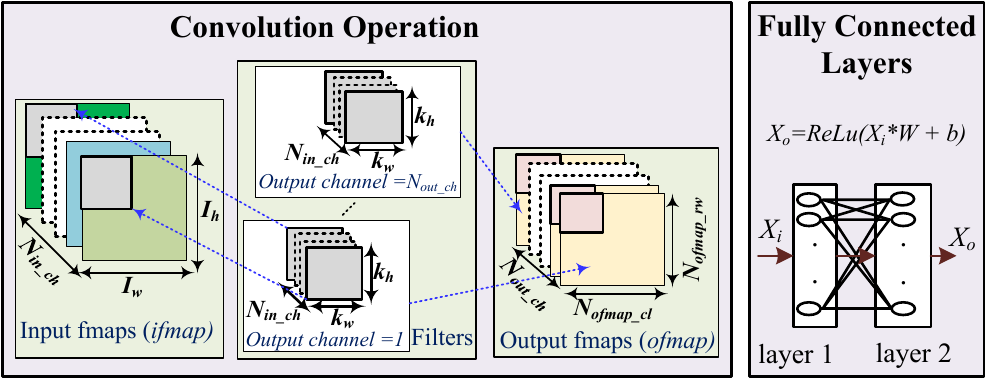}
	\vspace{-0.05in}
	\caption{Convolution and fully connected layer operations}
	\label{fig:conv_op}
	\vspace{-0.05in}
\end{figure} 

\begin{multline}
\scriptstyle
Ofmap[z][u][x][y] = ReLU(B[u] + \sum_{v=1}^{N_{in\_ch}} \sum_{i=1}^{k_{h}} \sum_{j=1}^{k_{w}} I[z][v][Sx+i][Sy+j]\\
\scriptstyle
\times F[u][v][i][j]), \\
\scriptstyle
0\leq z < N, 0\leq u < N_{out\_ch}, 0\leq y < N_{ofmap\_rw}, 0\leq x < N_{ofmap\_cl},\\
\scriptstyle
N_{ofmap\_rw}=((I_{h}-k_{h}+2P)/S) + 1,  N_{ofmap\_cl}=((I_{w}-k_{w}+2P)/S)+1
\end{multline}
where, Ofmap, $N$, $F$, $I$, $P$, $S$, and $B$ represent output feature maps, number of inputs in a Batch, filter weights, input feature maps, padding, stride size, and bias respectively \cite{eyeriss}.

Unlike the convolution layers,  each neuron of the Fully Connected (FC) layer is generally connected to every other neuron of its previous/next layer with a specific weight ($0$ for no connection) associated with each connection. The computations of FC layers are matrix/vector-matrix multiplications, where the output activation ($X_o$) of a layer is obtained by multiplying the input activation ($X_i$) matrix/vector with the weight matrix ($W$) followed by the addition of a bias term, and finally passing the result through a non-linear function such as ReLU,  $X_o=ReLu(X_i*W + b)$.

\subsection{Deep Learning/AI Hardware Accelerators}
SIMD, or Systolic array based hardware optimized for matrix (or vector)-matrix multiplication, is the present state-of-the-art hardware to accelerate AI operations \cite{aa1,tpu}.  The systolic array is only optimized for matrix-matrix multiplication, but it can not perform the dot product necessary for convolution layers. Mapping the convolution dot products into the matrix multiplications by converting the activation maps into the Toeplitz matrix and the kernel weights into a row vector is a popular solution to address this problem. Nonetheless, it involves redundant data in the input feature map which give rise to inefficient memory storage, and complex memory access pattern \cite{aa1}. More recently, heterogeneous architectures are evolving that have optimized cores for Convolution and FC layers \cite{DNPU}. While this solves the complications regarding Toeplitz matrix conversion, it incurs area overhead. Because when convolution core is active, FC core remains idle, wasting circuit area. In response to the existing issues, in this paper, we propose a novel concept of a reconfigurable core capable of efficiently performing both convolution dot products and matrix multiplications based on the operation-dependent (i.e., convolution or fully-connected) control signal. 

\subsection{Memory System in AI/Deep Learning Hardware}
The memory system is one of the vital metrics in determining the performance of AI hardware. Each off-chip DRAM access is 100 to 200 times more energy costly than any ALU operation or a local memory (e.g., register file/scratchpad) access \cite{aa1}. As a result, most energy-constraint AI hardware leverage a memory hierarchy of register file, global buffer, and DRAM. Moreover, a significant amount of memory is required to store the pre-trained weights for inference-only applications. The larger the global buffer memory, the more energy-efficient the AI hardware is due to lower DRAM access. Most of the existing DNN hardware use SRAM both as Global Buffer and Register file and eFlash as weight storage memory \cite{aa1}. Because of the large size of SRAM and static energy loss due to high leakage at scaled nodes (e.g., 10nm and newer), the global buffer size cannot be increased beyond a certain threshold energy-efficiently. eFlash starts to suffer from scaling challenges even at earlier technology such as 28nm \cite{samsung}. The benefits of our proposed $\Delta$-customized STT-MRAM as a replacement of both the SRAM-based global buffer and the eflash-based weight storage memory are many-fold: higher memory capacity, lower read-write latency and energy, and higher endurance against soft errors.

\section{Efficient AI/Deep Learning Hardware }
\begin{figure}[h]
	\centering
	\includegraphics[scale=0.67]{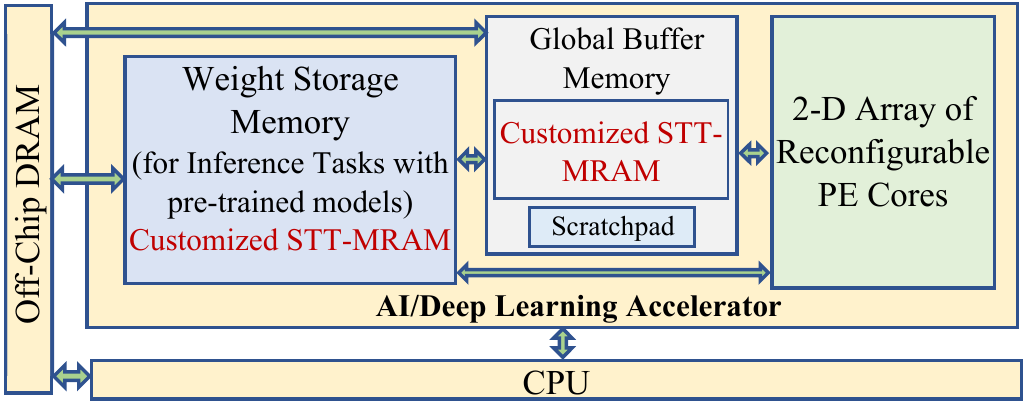}	
	\caption{AI Accelerator with reconfigurable cores optimized for both Conv. and FC layers, and STT-MRAM based on-chip memory.}
	\label{fig:arch}	
\end{figure}

Fig. \ref{fig:arch} depicts the top-level architecture of the accelerator containing the proposed reconfigurable core and MRAM-based memory system. The following sections describe the dataflow in convolution and systolic mode and formulate the memory occupancy time in each mode.
\subsection{Reconfigurable Core}
\begin{figure}[h]
	\centering
	\includegraphics[scale=0.7]{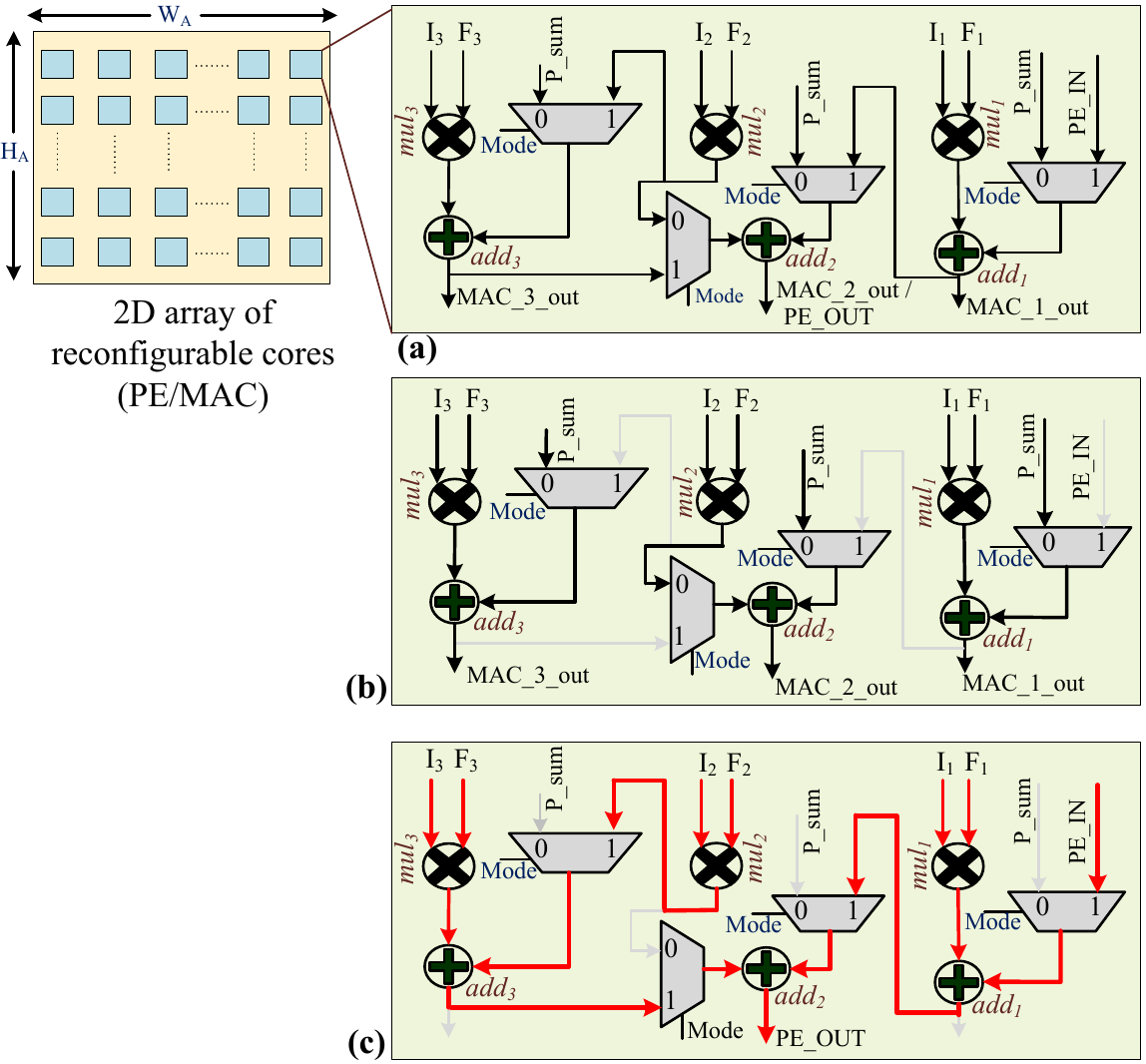}	
	\caption{(a) Reconfigurable core, (b) Reconfigurable core acting as building block of systolic array when mode is low, and (c) Reconfigrable core acting as convolution PE, when mode is high.}
	\label{fig:core}	
\end{figure}

The architecture and workflow of our proposed Reconfigurable Core are quite simple but powerful enough to support both matrix multiplication and convolution dot product at run-time configuration. The reconfigurable core consists of three MAC modules and four Multiplexers. Each MAC contains a BFloat16 multiplier and an FP32 adder \cite{bfloat,bf16_fb} to accommodate both training and inference. If only inference is desired, the hardware can be 8-bit int8 type \cite{aa1,tpu}. The multipliers take input feature maps and filter weights as inputs and pass the results to their neighboring adders to be added with the previous partial sum results. The multiplexers act as mode selectors of the core. When \textit{Mode} is de-asserted, the MACs are disconnected from each other and their outputs are collected downward to reflect the systolic array architecture (Fig. \ref{fig:core}(b)). On the other hand, when \textit{Mode} is asserted, three MACs collectively act as a convolution block that performs three dot products parallelly and produces one partial sum (Fig. \ref{fig:core}(c)). In this case, $adder_{3}$ adds the outputs of $multiplier_{3}$ and $multiplier_{2}$ to produce the intermediate sum. Meanwhile, $adder_{1}$ adds the $multipler_{1}$ output with the previous partial sum. These operations occur concurrently, provided that the input activations and filter weights are assigned to the multipliers parallelly. Once the outputs from $adder_{3}$ and $adder_{1}$ are ready, $adder_{2}$ sums them up to produce the PE_OUT. The building block containing three MACs and four Muxes is defined as a Process Element (PE) block for convolution in this work. Fig. \ref{fig:core} illustrates the functionality of the Reconfigurable core in systolic array mode (b) and convolution mode (c).

\subsection{STT-MRAM Based On-Chip Memory System}
The prime criteria of memory to sustain as an on-chip memory are: high density, low read/write latency and  energy. Conventional STT-MRAM suffers from high read/write energy and latency. However, in the case of on-chip memory/global buffer, the intrinsic non-volatility property of STT-MARM can be compromised to minimize the read/write energy and latency by adjusting the thermal stability factor ($\Delta$). Considering the data retention time in the global buffer, $\Delta$ can be scaled down to achieve a significant reduction in read/write energy,  latency, and increase in cell density. This subsection will formulate necessary expressions to calculate the data retention time in the global buffer for the most time-consuming AI operations, such as the convolution layer and fully connected layer operations. The derived expressions will help us to precisely determine the maximum data retention time in global buffer, and thus help to scale down $\Delta$. 

Deep Learning/AI operations are layer-wise sequential operations, meaning the current layer's output acts as input to the following layer. To formulate the data retention time between two consecutive layers, in inference mode, we define \textbf{$T_{1}$} as the time required by the accelerator  to generate the ofmap of one layer. Once the ofmap of one layer is generated, it goes through Maxpooling and Activation functions (e.g., ReLU) to serve as the input to the following layer. We refer \textbf{$T_{pool\_relu}$} as the time required to perform the Maxpooling and ReLU operations. The time required to generate the ofmap of the following layer is termed as \textbf{$T_{2}$}. Finally, \textbf{$T_{ret}$} is the data retention time in memory between two consecutive convolution (or fully connected) layers.

\subsubsection{Retention time for Conv-Conv layers}
In convolution mode, each PE block of the array performs the dot product between the input feature maps (ifmaps) and weights. Each unit PE block's size is defined as $P_s$, where $P_s$ represents the number of elements the MAC module can process. The ifmaps and kernel weights are loaded into the PE array from global buffer memory, and the PE array computations occur in parallel. Without loss of generality, in our analysis, we adopted the Row Stationary data flow where kernel rows are loaded into the PE blocks and kept stationary, and ifmaps are loaded and shifted according to the $stride$ size \cite{aa1,eyeriss}. The partial sums are accumulated vertically to generate the output feature maps (ofmaps). This process is repeated until a complete $ofmap$ is generated. Setting $Mode=1$ in the Muxes of the PE blocks (Fig. \ref{fig:core}) ensures that the Reconfigurable core is acting as a Convolution core.

To calculate $T_1$, we formulate an expression that helps us estimate the time required to generate the output (ofmap) of a convolution layer. We assume that the operations related to the next output channel will be assigned in the accelerator array only after all the MAC operations related to the previous output channel have been completed. In other words, in an iteration of the accelerator array, the input channels present in it are all related to the same output channel. In addition to simplifying the PE scheduling procedure, this assumption also aligns with our goal of obtaining a convolution layer's worst-case completion time.

\begin{figure}[h]
	\centering
	\includegraphics[width=3.2in,height=2in]{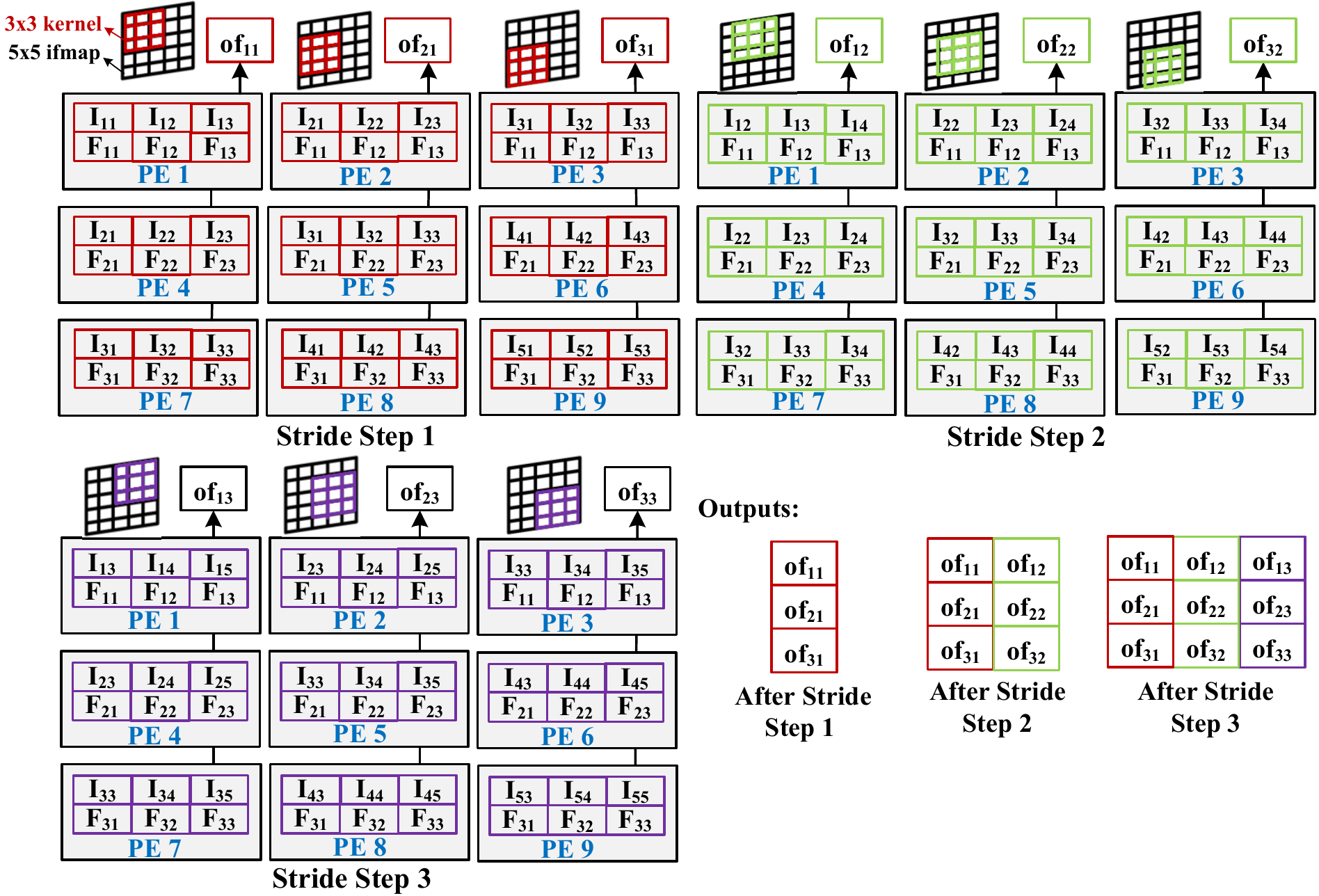}
	\caption{A 3\texttimes3 kernel ($k_h=k_w=3$) is convolved (with stride=1) over a 5\texttimes 5 ifmap to produce 3\texttimes 3 ofmap ($N_{ofmp\_rw}=N_{ofmp\_cl}=3$). The size of unit PE block, $P_s = 3$. Total 9 PE blocks are required for this convolution.}
	\label{fig:conv}
\end{figure}

For layer $n-1$, a single row of a partial ofmap (i.e., ofmap corresponding to one kernel and one input channel) will require, ($k_h*\ceil*{k_w/P_s})$ PE blocks (symbol meanings are given in Table 1), implying that a partial ofmap for a single input channel will require, $N_{ofmp\_rw}*k_h*\ceil*{k_w/P_s}$ PEs. (The $\ceil*{}$ symbol means $ceil$ operation where the result is rounded to the nearerst larger integer). An example of convolution operation inside the core in Conv. mode is shown in Fig. \ref{fig:conv}.

\begin{table}[]
\centering
\scriptsize
\caption{Parameters \& description}
\vspace{-0.05in}
\renewcommand{\arraystretch}{1.1}
\begin{tabular}{|c|c|}
\hline
Parameter     & Description                                                                      \\ \hline
$N_{in\_ch}$            & Number of input channels                                                                 \\ \hline
$N_{out\_chn}$            & Number of input channels                                                                 \\ \hline
$N_{bat}$            & Number of images per mini batch                                                                 \\ \hline
$k_h$                  & kernel height                                                                            \\ \hline
$k_w$                  & kernel width                                                                             \\ \hline
$P_s$                  & PE internal size                                                                         \\ \hline
$W_A$                  & Accelerator array width (PEs)                                                                  \\ \hline
$H_A$                  & Accelerator array height (PEs)                                                                 \\ \hline
$W_{SA}$            & Systolic array width (\# of MACs), \; $P_s*W_A$                                                                 \\ \hline
$N_{ofmap\_rw}$         & Number of rows in ofmap                                                                  \\ \hline
$N_{ofmap\_cl}$         & Number of columns in ofmap                                                               \\ \hline
$N_{cyc\_per\_stp}$ & \begin{tabular}[c]{@{}c@{}}Total clk cycles per iteration in conv./systolic mode of accelerator\end{tabular} \\ \hline
$n_{fc}$                  & number of neurons in input FC layer                                                                           \\ \hline
$m_{fc}$                  & number of neurons in output FC layer                                                                         \\ \hline
$T_{clk}$                  & Clock cycle time                                                                         \\ \hline
\end{tabular}
\end{table}

Next, we find out how many partial ofmaps (i.e., how many input channels) can be fitted in the full accelerator array in one step. This number is obtained by dividing the total available PEs in the accelerator array, $W_A*H_A$, by the number of PEs required for one input channel. The total number of required steps (i.e, number of times the complete accelerator array will be used) for all input channels ($N_{in\_ch}$) for one 3D filter (i.e, one output channel), $N_{steps\_per\_out\_ch}$, can be expressed  as,

\begin{equation}
\scriptstyle
N_{steps\_per\_out\_ch}=\ceil*{\frac{N_{in\_ch}\;*\;k_h\;*\;N_{ofmp\_rw}\;*\;\ceil*{\frac{k_w}{P_{s}}}}{W_A\;*\;H_{A}}}
\label{eq:step_out_ch}	
\end{equation}

The details of the symbols used in Equations (2)-(6) can be found in Table 1. Inside the accelerator, time for each of the above steps,

\begin{equation}
t_{per\_step}=T_{clk}\;*\;N_{cyc\_per\_stp}\;*\;N_{ofmp\_cl}\;*\;N_{bat} 
\label{eq:t_per_stp}	
\end{equation}

$N_{cyc\_per\_stp}$ refers to the total clock cycles required in the accelerator, for one image of the batch, to perform, (i) dot products between the kernel and ifmap elements, (ii) partial sum accumulation of the dot products, and (iii) partial sum of ofmap of previous input channel with current channel. This term depends on the circuit-level implementation of accelerator hardware. The term $N_{ofmap\_cl}$ appears in Equation (\ref{eq:t_per_stp}) because the kernel needs to be shifted (i.e., according to the $stride$ parameter of convolution) this many times to generate the partial ofmap for each input channel. $N_{bat}$ appears in the equation since each image from the mini-batch will be serially processed \cite{eyeriss}. In between each input channel operations, the partial sum of ofmaps of the input channels will be stored in the scratchpad to be accumulated to the next input channel's partial ofmaps to finally create the full ofmap output for that particular output channel and filter. The total time required to generate each output channel/ofmap, $t_{per\_out\_ch}$, is given by,

\begin{equation}
t_{per\_out\_ch}=N_{steps\_per\_out\_ch}\;*\;t_{per\_step}
\label{eq:t_per_out_ch}
\end{equation}

If there are a total of $N_{out\_chn}$ output channels, then the total time required to generate the full ofmap (i.e., for all output channels) is, $T_1=t_{per\_out\_ch}*N_{out\_chn}$. Using Equations (\ref{eq:step_out_ch})-(\ref{eq:t_per_out_ch}), the $T_1$ term can be expressed with the following equation. All parameters in Equation (\ref{eq:conv_Exp}) are for Conv. layer $n-1$.

\begin{equation}
\scriptstyle
    T_1=\scriptstyle \ceil*{\frac{N_{in\_ch}\;*\;k_h\;*\;N_{ofmp\_rw}*\ceil*{\frac{k_w}{ P_{s}}}}{W_A\;*\;H_{A}}}*\;\scriptstyle T_{clk}*\;N_{cyc\_per\_stp}\;*\;N_{ofmp\_cl}\;*\;N_{bat}*N_{out\_chn}
\label{eq:conv_Exp}
\end{equation}

The ofmap of layer $n-1$  will act as ifmap to next layer $n$ after passing through the ReLU and MaxPool layers. The $ifmap_n$ should be retained in the memory until layer $n$ has finished reading it to generate its output $ofmap_n$. A closer look into the situation will reveal that the input data read time for layer $n$ is related to the $ofmap_n$ generation time. 
This implies that the $ifmap_n$ data need to be in the memory for a maximum duration of time that is equal to the time required for complete $ofmap_n$ generation. Considering the above facts, we first calculate  $ofmap_n$ generation time using the similar methods of Equations (2)-(\ref{eq:conv_Exp}) and then assign it as $T_2$. All parameters in Equation (\ref{eq:conv_T2}) are for Conv. layer $n$.

\begin{equation}
\scriptstyle
    T_2=\scriptstyle \ceil*{\frac{N_{in\_ch}\;*\;k_h\;*\;N_{ofmp\_rw}*\ceil*{\frac{k_w}{ P_{s}}}}{W_A\;*\;H_{A}}}*\;\scriptstyle T_{clk}*\;N_{cyc\_per\_stp}\;*\;N_{ofmp\_cl}\;*\;N_{bat}*N_{out\_chn}
	\label{eq:conv_T2}
\end{equation}

ReLU and MaxPool layers take relatively much shorter time and also do not  involve complex computations as Conv. layers do. Therefore, we can directly estimate $T_{pool\_relu}$ from hardware implementation of ReLU and MaxPool layers. Combining $T_1$, $T_2$, and $T_{pool\_relu}$, we can estimate the required data retention time, $T_{ret}$, in memory between two consecutive Conv. layers in inference phase.

\begin{equation}
T_{ret\_conv-conv} = T\textsubscript{1} + T\textsubscript{pool_relu} + T\textsubscript{2}
\label{eq:conv_ret_time}
\end{equation}

\begin{figure}[h]
	\centering
	\includegraphics[scale=.65]{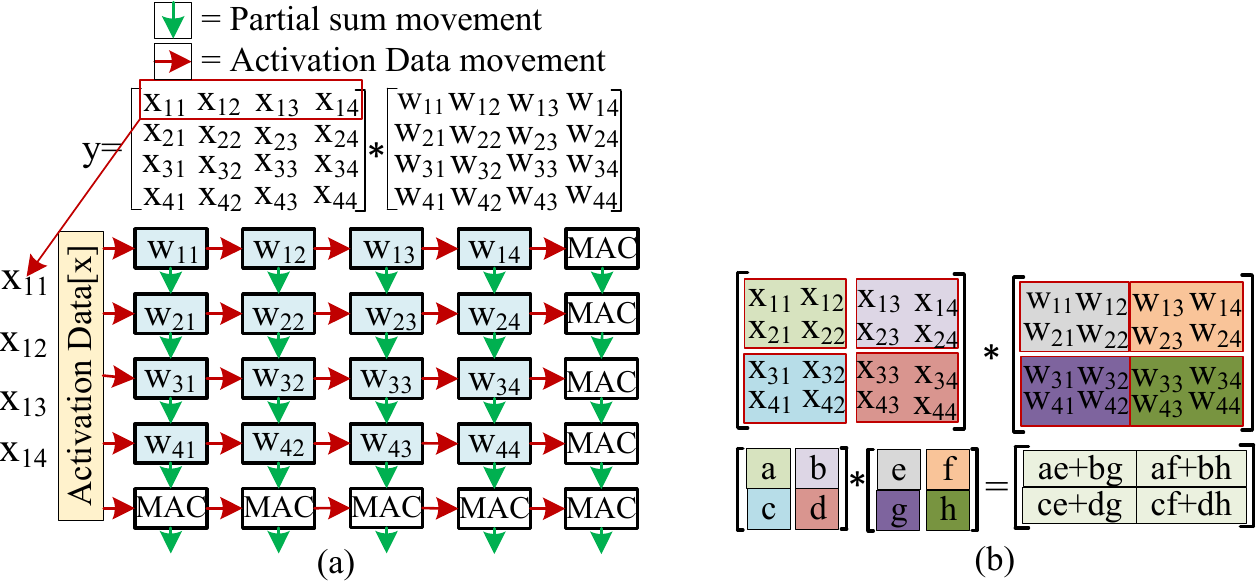}
	\vspace{-0.05in}
	\caption{(a) Dataflow inside systolic array, (b) Larger matrices can be divided into smaller sub-matrices to fit in the systolic array. An example of dividing two 4\texttimes4 matrix into four 2\texttimes2 matrices to fit into 2\texttimes2 systolic array.}
	\label{fig:systolic_df}
	\vspace{-0.05in}
\end{figure}

\subsubsection{Retention time for FC-FC layers}
To perform the computations associated with FC layers, the Reconfigurable Core transforms into the Systolic array. 
This is achieved by disabling the Mode signal of the Muxes present in the PE blocks. In this section, we formulate an expression to estimate the time required to compute the output of an FC layer. The systolic array shown in Fig. \ref{fig:core} (b) has $H_{A} \times W_{SA}$ MAC modules. Because of the reconfigurable feature of the core, $W_{SA} = P_{s} * W_{A}$. The weights are loaded into the array according to the capacity of the systolic array implying that the number of weights can be loaded into the array in one step is equal to the number of MACs present in the array, $N_{wt\_per\_step} = H_{A}*W_{SA}$. If the total number of weights is greater than the number of weights the array can accommodate in one step, i.e., $N_{tot\_wt} > N_{wt\_per\_step}$, using the concept of \textit{divide \& conquer} in matrix multiplication (Fig. \ref{fig:systolic_df} (b)) we find out how many steps (i.e., number of times we need to load new weights to the accelerator array) are required to complete the computation with all elements of the weight matrix. The number of steps required to complete the computation with all weights is, $\ceil*{m_{fc} / H_{A}}\;*\;\ceil*{n_{fc} / W_{SA}}$. (For symbol meanings see Table 1). In every step, the array is loaded with $N_{wt\_per\_step}$ weights, inputs are streamed from left to right, and the partial sums move downward to be collected in accumulators \cite{tpu}. The clock cycles required to complete each step are $N_{cyc\_per\_stp}$ that depends on the circuit-level implementation of systolic array hardware and the dimension of systolic array core. Combining the above terms, and considering there are $N_{bat}$ images in the mini-batch, time required to generate the output of FC layer $(n-1)$, $T_1$, is expressed in Equation (\ref{eq:T1_fc_fc}), where all parameters are for FC layer $(n-1)$.


\begin{equation}
\scriptstyle
  T_1 = \scriptstyle \ceil*{\frac{m_{fc}}{H_{A}}}\;*\;\ceil*{\frac{n_{fc}}{W_{SA}}}\;*\;T_{clk}\;*\;N_{cyc\_per\_stp}\;*\;N_{bat}
  \label{eq:T1_fc_fc}
\end{equation}

FC layer $n$ will consider the output of previous FC layer $(n-1)$ as its input. The output of FC\textsubscript{n-1} should be stored in the memory until FC\textsubscript{n} has completed reading it for generating its output. With this reasoning, we calculate the output generation time for FC\textsubscript{n} following the above method and assign it as $T_2$. All parameters of Equation (\ref{eq:T2_fc_fc}) are for FC layer $n$. 


\begin{equation}
\scriptstyle
  T_2 = \scriptstyle \ceil*{\frac{m_{fc}}{H_{A}}}\;*\;\ceil*{\frac{n_{fc}}{W_{SA}}}\;*\;T_{clk}\;*\;N_{cyc\_per\_stp}\;*\;N_{bat}
\label{eq:T2_fc_fc}	
\end{equation}

Two consecutive FC layers do not have MaxPooling layer in between. Therefore, we can find the data retention time, $T_{ret\_fc-fc}$,  for an FC layer followed by another FC as, 
\begin{equation}
T_{ret\_fc-fc} = T_{1}+T_{2}
\label{eq:Tret_fc_fc}
\end{equation}

\subsection{Retention time of Convolution layer followed by FC layer} The retention time between a Convolution layer followed by an FC layer is also expressed as,
\begin{equation}
T_{ret\_conv-fc}\; = T_1\;+T_{pool\_relu}\;+T_2 
\label{eq:Tret_conv_fc}
\end{equation}
Here, T\textsubscript{1} is the time required to generate the Conv. layer ofmap and T\textsubscript{2} is the time required to generate the FC layer output. 


Using the above expressions of weight, ifmap, and ofmap occupancy times in global buffer memory, for a particular accelerator hardware architecture and the operating clock frequency, we can estimate the maximum retention time required for STT-MRAM based global buffers. The MRAM Write and Read times will be added with the above retention time expressions. As the MRAM Read/Write times are orders of magnitude lower (i.e., less than 10ns) compared to the retention times $T_1$ and $T_2$ which are in the $ms$ or $s$ range as explained in Section IV, we did not explicitly add the MRAM Read/Write time with the above retention time ($T_{ret}$) expressions.

\section{Optimizing STT-MRAM for AI Accelerators }

\label{sec:MRAM basics}
\begin{figure}[H]
	\centering
	\includegraphics[scale=0.6]{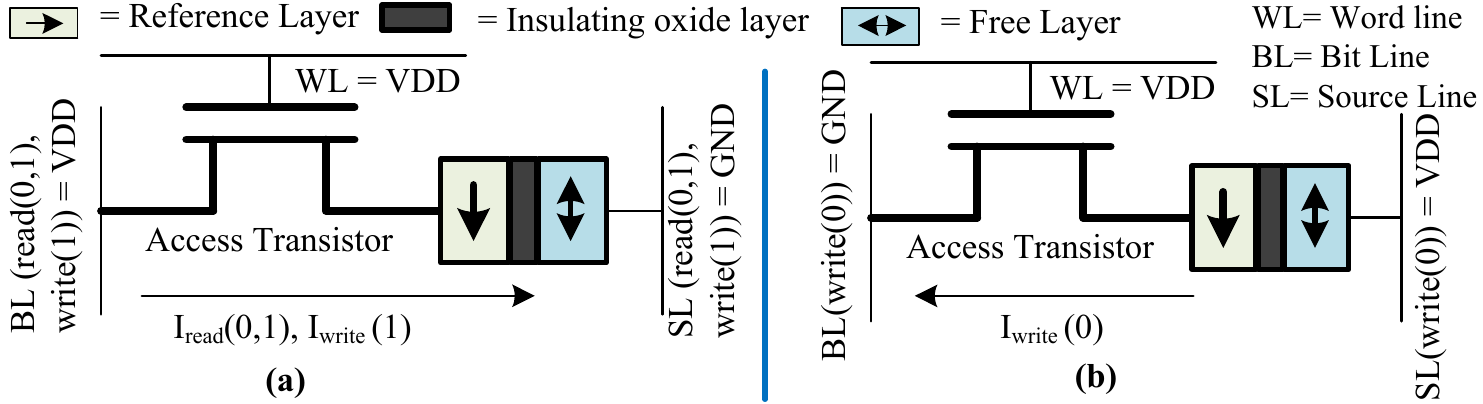}
	\vspace{-0.02in}
	\caption{Bit cell of STT-MRAM. (a) shows reading from it and writing 1, (b) shows writing 0.}
	\label{stt-fig}
	\vspace{-0.02in}
\end{figure}

A bit cell of STT-MRAM consists of a Magnetic Tunnel Junction (MTJ) for storing the bit and an access transistor to read/write the bit. The MTJ contains two ferromagnetic layers, one with fixed magnetic orientation, and another free layer whose orientation can be switched externally by an applied current. The orientation of the free layer relative to the reference layer represents the state of the stored bit; parallel orientation refers to logic 0, while anti-parallel orientation refers to logic 1. Fig. \ref{stt-fig} depicts the cell schematic, read, and write operations.

\subsection{Critical Design and Performance Parameters of MTJ}
\subsubsection{Thermal Stability Factor and Critical Current} The energy barrier $E_b$, which the free layer magnetization must overcome to switch its stable state is defined as the thermal stability factor ($\Delta$), and is expressed as \cite{stt_eqn1,stt_eqn2}:
\begin{equation}
\vspace{-0.02in}
\Delta = \frac{E_{b}}{k_{B}T} = \frac{H_{K}M_{S}V}{2k_{B}T}
\label{eq:delta}
\vspace{-0.02in}
\end{equation}
Where, $E_{b}$ = Energy Barrier of free layer, $k_{B}$ = Boltzmann Constant, $T$ = Temperature, $H_{K}$ = Anisotropy field, $M_{S}$ = Saturation Demagnetization, $V$ = Volume of MTJ. 

Critical current, $I_{c}$, is defined as the minimum current required to flip the state of the free layer \cite{stt_eqn1,stt_eqn2,stt_model}. The critical switching current is modeled as\cite{stt_eqn1,stt_eqn2}:
\begin{equation}
\vspace{-0.02in}
I_{c}=(\frac{4ek_{B}T}{h})*\frac{\alpha}{\eta}*\Delta*(1+\frac{4\pi M_{eff}}{2H_{K}})
\label{eq:ic}
\vspace{-0.02in}
\end{equation} 
Where, $e$ = electron charge, $k_{B}$ = Boltzmann Constant, $T$ = Temperature, $h$ = Plank's Constant, $\alpha$ = LLGE damping constant, $\eta$ = STT-MRAM efficiency parameter, $4\pi M_{eff}$ = Effective demagnetization field, and $H_{K}$ = Anisotropy field. 

\subsubsection{Retention Time \& Retention Failure} Once data is written, MTJ should retain its state, even if the power source is removed, until any external force is applied to flip the state. However, due to thermal noise, the logic state might get flipped unintentionally. The maximum time MTJ can retain its non-volatility is known as data retention time. The retention failure probability for a given time period $t_{ret}$ is \cite{stt_eqn1,stt_eqn2}: 
\begin{equation}
\vspace{-0.02in}
P_{RF} = 1- exp \left[-\frac{t_{ret}}{\tau* exp(\Delta)} \right]
\label{eq:ret}
\vspace{-0.02in}
\end{equation}
Where, $t_{ret}$= retention time, $\tau$= technology constant.

\subsubsection{Read Disturbance (RD)} To read a bit from STT-MRAM, read current $I_{r}$, much less than the critical current $I_{c}$, is flown from bit line through the access transistor and MTJ. Fig. \ref{stt-fig} shows that writing 1 and reading (both 0 \& 1) share the same current path. This can sometimes cause the unintentional switching of the bit-cell content resulting in Read Disturbance (RD). For read current $I_{r}$ and read latency $t_{r}$, the probability of RD can be modeled as\cite{stt_eqn1,stt_eqn2}:
\begin{equation}
\vspace{-0.02in}
P_{RD} = 1 - exp \left[-\frac{t_r}{\tau * exp(\Delta (1- \frac{I_r}{I_c}))} \right]
\label{eq:rd}
\vspace{-0.02in}
\end{equation}

\subsubsection{Write Error Rate (WER)} Writing a bit cell requires a write current $I_{w}$,  larger than $ I_{c}$, to be flown between BL to SL as shown in Fig. \ref{stt-fig}. Because of the stochastic nature of the write operation, the switching time of MTJ varies from access to access \cite{stt_eqn1,stt_eqn2,stt_model}. If the write current is terminated before the free layer has successfully changed its state, the write operation can be erroneous. For write pulse width $t_w$, the Write Error Rate (WER) is \cite{stt_eqn1,stt_eqn2}: 
\begin{equation}
\vspace{-0.02in}
WER_{bit} = 1- exp \left[\frac{-\pi ^{2}\Delta(\frac{I_{w}}{I_{c}}-1)}{4[\frac{I_w}{I_c}*exp\{\frac{t_w}{\tau}(\frac{I_w}{I_c}-1)\}-1]} \right]
\label{eq:wer}
\vspace{-0.02in}
\end{equation}

\subsection{Customizing STT-MRAM For AI Accelerators}
\subsubsection{Scaling Thermal Stability Factor} To achieve typical retention period of 10 years, the thermal stability factor, $\Delta \geq 60$ is required \cite{scale_thermal,intel,samsung,stt_eqn1,stt_eqn2,stt_model}. However, such a long retention time may be unnecessary depending on the application. For example, if MRAM is used as the global buffer memory in AI accelerators, then the retention time can be significantly scaled depending on the weight and input/output feature-map data occupancy time (e.g., $ms$ to $s$ range) in that memory. If STT-MRAM is used as eFlash replacement for pre-trained weight storage for AI inference tasks, then 3 to 5 years retention might be enough instead of 10 years. From Equation (\ref{eq:delta}), it is seen that by adjusting the volume (i.e., area and/or thickness) of the MTJ the thermal stability factor ($\Delta$) can be scaled. In other words, considering the target operating temperature range of the AI accelerator and the expected life-time of the data, scaling down of thermal stability factor will improve area efficiency by increasing the memory bit-cell density. Moreover, with scaled $\Delta$ and bit-cell area, the cell would require a lower operating current, thus saving energy.

\subsubsection{Optimizing Read/Write Latency and Energy at Target WER and RD} Recent state-of-the-art STT-MRAMs can compete or outperform SRAMs in all aspects except write energy and write latency \cite{imec,intel,stt_ram_tsmc2}. However, for AI accelerator applications, by scaling $\Delta$ and the retention time of STT-MRAM we can circumvent the write energy and latency limitations.  Equation (\ref{eq:wer}) implies that write latency, $t_{pw} \, \propto \, ln(\Delta)$ at constant write error rate. We can exploit this relationship to reduce the write latency with scaling down of $\Delta$.  From Equation (\ref{eq:ret}) we infer that retention time $t_{ret}$ is exponentially proportional to $\Delta$. Thus, depending on the desired retention period of STT-MRAM in AI accelerator, we can optimally scale down $\Delta$, and also minimize write latency  at that target retention time. However, Equation (\ref{eq:wer}) also implies an inverse relationship between write latency and write error rate, which hinders us from aggressively scaling down write latency at the desired $\Delta$. Fortunately, to boost the writing speed at the scaled $\Delta$, we can keep $I_{w}$ higher (e.g., close to the pre-scaled value), and this can assist in designing a STT-MRAM with high write-speed \cite{stt_model}. Recently, high-speed write has been experimentally demonstrated in \cite{ibm_2019} by optimizing the free layer materials. We can identify the optimum $\Delta$ and $I_{w}$ that minimizes write latency, write energy while still satisfying the WER and retention time requirements for the AI accelerator. As depicted in Equation (\ref{eq:ic}), with scaling of $\Delta$ the critical current $I_{c}$ decreases linearly, and hence read current $I_{r}$ also decreases. At this scaled $\Delta$ and $I_{r}$, the read latency can also be scaled by adjusting the sense amplifier reference voltage \cite{stt_model,imec}. Equation (\ref{eq:rd}) implies that at scaled $\Delta$, the shortened read pulse duration will also ensure that the Read Disturb rate is within the acceptable target.

\subsection{Addressing Process and Temperature Variation}
The performance of MRAM can degrade due to the process and temperature variations \cite{intel,imec,tdk_18}. Process-induced variations in free layer thickness in MTJ, and in access transistor channel length/width and threshold voltage contribute to the performance variations in MRAM. From Silicon measurement data in \cite{imec}, the standard deviation ($\sigma$) of MTJ diameter variation  was reported to be 2.1\% of the mean. Magnetic Anisotropy field ($H_{K}$) is another source of process variation in STT-MRAM.

\begin{figure}[H]
	\centering
	\includegraphics[scale=0.8]{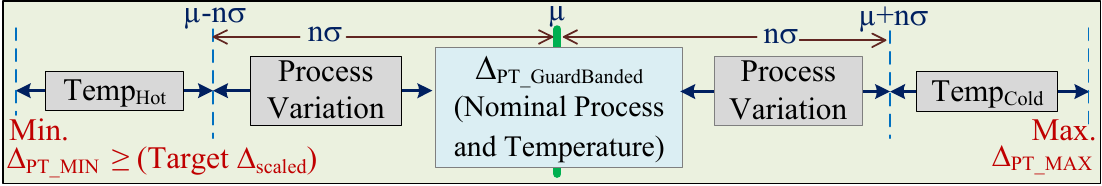}
	\caption{Impact of process and temperature variation on thermal stability factor ($\Delta$).}
	\label{fig:delta_var}
\end{figure}

The bit-cells are placed compactly on the layout, as a result, the bit-cell to bit-cell variations within the same die/chip are minimal, and the process variation is dominated by the chip-to-chip variations. $\Delta$ increases with an increase in MTJ diameter and $H_{K}$ due to process variation, and a decrease in temperature from the nominal value (Fig. \ref{fig:delta_var} and Equation \ref{eq:delta}). An increase in $\Delta$ increases critical current ($I_c$) which eventually increases the write current ($I_w$) (Equation \ref{eq:ic} and \ref{eq:wer}). Given the smaller write pulse, write failure occurs when supplied $I_w$ is less than the required $I_w$. Worst-case occurs when both, (i) the supplied $I_w$ decreases due to the access transistor being in the slow process corner, and (ii) the required $I_w$ increases due to increase in $\Delta$ resulting from Process and runtime Temperature (PT) variations. On the other hand, decrease in $\Delta$ beyond a minimum due to PT variation will result in retention failure (Equation \ref{eq:ret}).

\begin{figure}[H]
	\centering
	\includegraphics[width=3.4in,height=1in]{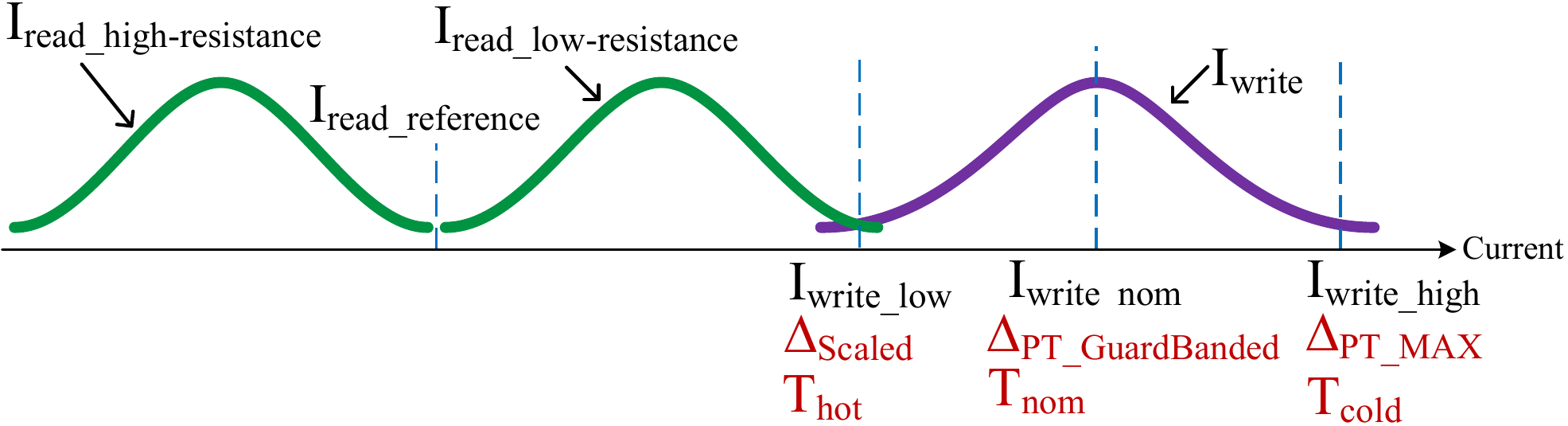}
	\caption{Distribution of read/write currents with process variation. Worst-case occurs when worst process corners experience $T_{hot}$ or $T_{cold}$.}
	\label{fig:variation}
\end{figure}

To protect the desired $\Delta_{scaled}$ against the worst-case PT variation,  appropriate Guard-Band needs to be added. The $\Delta_{PT\_GuardBanded}$ is chosen to cover both the worst-case 4$\sigma$ range (i.e., 99.993\% of the samples) of process variation and high temperature operating scenario as shown in Equation \ref{eq:delta_min}.
\begin{equation}
\Delta_{scaled}\leq(\Delta_{PT\_GuardBanded}-4\sigma)*(T_{nom}/T_{hot})
\label{eq:delta_min}	
\end{equation}

\begin{equation}
\Delta_{PT\_MAX}=(\Delta_{PT\_GuardBanded}+4\sigma)*(T_{nom}/T_{cold})
\label{eq:delta_max}	
\end{equation}

\begin{figure}[H]
	\centering
	\includegraphics[scale=0.7]{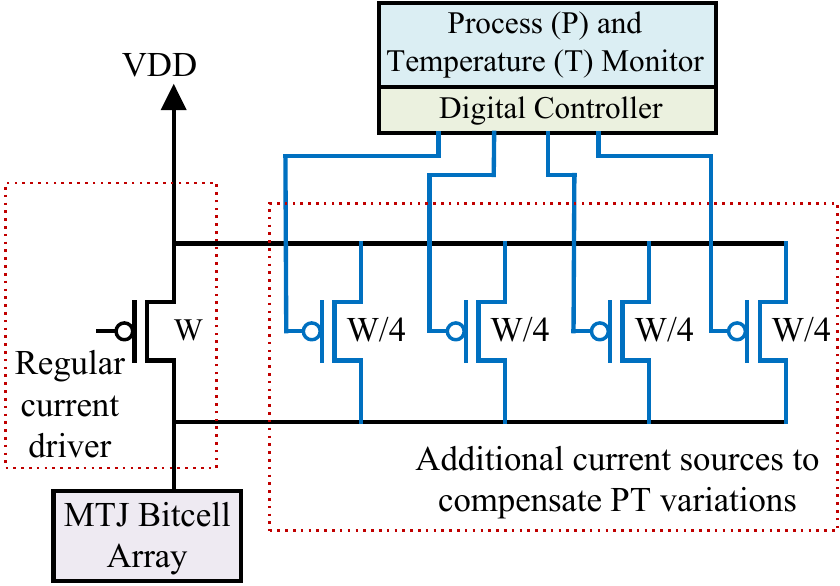}
	\caption{Modified Write Driver}
	\label{fig:write_driver}
\end{figure}

The chip samples located on the right side of the process variation distribution (i.e., $\mu +n*\sigma$, where $n\geq1$) of $\Delta_{PT\_GuardBanded}$,  will experience larger $\Delta$ as shown in Fig. \ref{fig:delta_var} and \ref{fig:variation}. Additionally, at cold temperatures, the $\Delta$ will further  increase to $\Delta_{PT\_MAX}$ as shown in Equation \ref{eq:delta_max}. Although the higher $\Delta_{PT\_MAX}>\Delta_{scaled}$ will be benign for retention time, the required write current will increase in this scenario to confine the write time and Write Error Rate (WER) of these samples within the nominal bound. Designing the write driver for this worst-case scenario will dissipate unnecessary power for all other non-worst-case samples. To address this, we propose a dynamically adjustable write driver depicted in Fig. \ref{fig:write_driver}. The proposed write-driver circuit provides additional write current in extreme PT conditions. The Process and Temperature Monitor (PTM) block continuously monitors the process and temperature changes. In addition to adjusting the write current according to process variation profile of the MRAM chip/die, the PTM also senses the runtime temperature and adjusts the current dynamically by turning on/off the additional PMOS transistors (Fig. \ref{fig:write_driver}). For example, at the nominal process and temperature, the extra transistors can be off, however, at PT-induced rise in $\Delta$ the transistors can be individually activated to ensure successful write. 

In summary, we design the MTJ with higher $\Delta_{PT\_GuardBanded}$ than the desired $\Delta_{scaled}$ to accommodate potential degradation in thermal stability from worst-case 4$\sigma$ process variation and runtime high temperature, and proposed a write-driver with controllable current drive to address the high write-current demand at cold temperature and slow process corner.

\subsection{MRAM Write Energy Optimization in Accelerator with ScratchPad}
Addressing the fact that the write energy of STT-MRAM is higher than the read energy, we propose an innovative scratchpad-assisted MRAM global buffer architecture to minimize the write frequency in STT-MRAM, and thus further optimize the energy. Reduced write-frequency is achieved by using a small global SRAM scratchpad, typically in the KB range (details in Section V), in addition to a large (i.e., MB range) global STT-MRAM buffer. When the accelerator PE array generates partial ofmaps (i.e., ofmap corresponding to each input channel), they need to be stored somewhere in memory to be added to the next partial ofmap to produce the complete ofmap of an output channel. The reason behind these partial-ofmap writes is that the accelerator might not produce the complete ofmap in one step. Between the subsequent steps, the partial ofmap result from the previous step needs to be written in the memory to be subsequently read and accumulated with the partial ofmap result from the following step. Adding this small SRAM scratchpad memory (for intermediate ofmap writes) with MRAM global buffer further improves the energy efficiency of MRAM-buffer-based deep learning accelerators. In summary, our proposed scratchpad-assisted MRAM memory architecture provides energy efficiency by, (i) minimizing the STT-MRAM write frequency, and (ii) additionally, at a smaller size, SRAM is more energy-efficient than STT-MRAM \cite{imec}.
\section{Results and Analysis}
\vspace{-0.01in}
\label{sec:Retention Time}

\subsection{Design Space Exploration for Selecting Memory Capacity}
\begin{figure}[h!]
	\centering
	\includegraphics[scale=0.97]{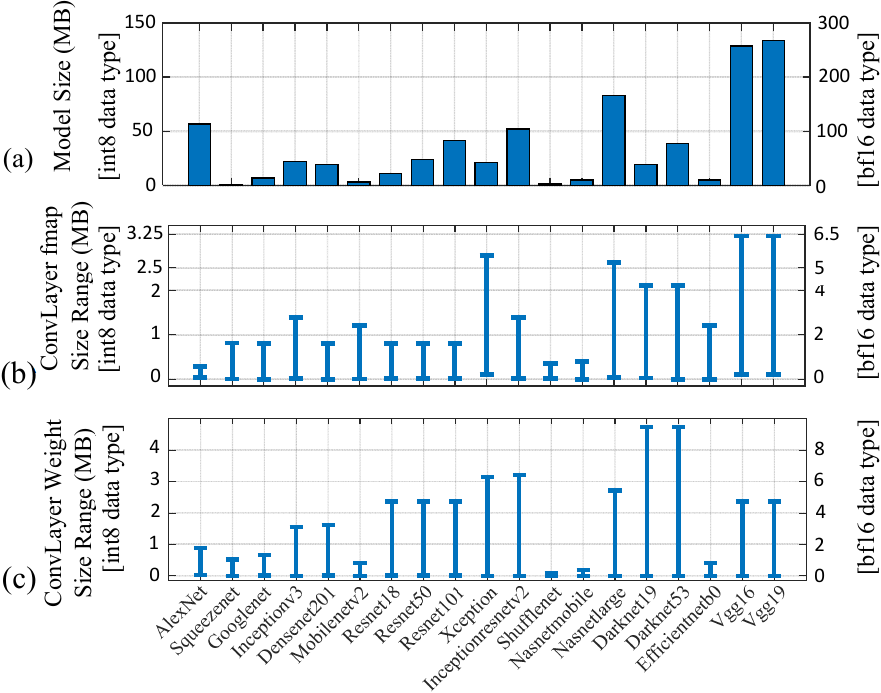}
	\caption {(a) Complete sizes of widely used AI models. (b) Activation map (ofmap/ifmap) sizes,  (c) Weight sizes for Conv layers. }
	\label{fig:model_size}
\end{figure}

Nineteen widely used state-of-the-art deep learning models were analyzed to design and validate our STT-MRAM based AI accelerator with the reconfigurable core. Fig. \ref{fig:model_size} (a) shows the model sizes both in 8-bit int8 (left Y-axis) and 16-bit BrainFloat16 (BF16) \cite{bfloat,bf16_fb} (right Y-axis) datatypes. For inference-only accelerator int8 datatype and hardware suffice, however, if full-scale training or transfer-learning is desired then BF16 hardware and data-type are necessary \cite{bfloat,bf16_fb}. The models' sizes imply that around 280MB and 140MB of STT-MRAM is required as non-volatile (NVM) weight storage memory to store the pre-trained models using BF16 and int8  datatypes, respectively. The STT-MARM non-volatile weight storage memory can replace the currently used eFlash memory as an efficient alternative. Fig. \ref{fig:model_size} (b) and (c) represent the input/output featuremap and weight size ranges of all models for convolution layers both in int8 (left Y-axis) and BF16 (right Y-axis) formats, and these data helps us to estimate the maximum required global buffer (GLB) memory size to avoid DRAM accesses during each convolution layer operation. In the cases of fully-connected layer operations, only the featuremaps, usually in KB range for most of the models, are stored in the GLB, and the weights, around 200MB in size for the largest model in BF16, are directly assigned from DRAM (or weight-storage NVM) to the systolic array for matrix multiplications. Hence, we ignored the fully connected layers' weight and activation sizes from design space analysis for selecting on-chip GLB memory capacity.

\begin{figure}[h!]
	\centering
	\includegraphics[scale=0.94]{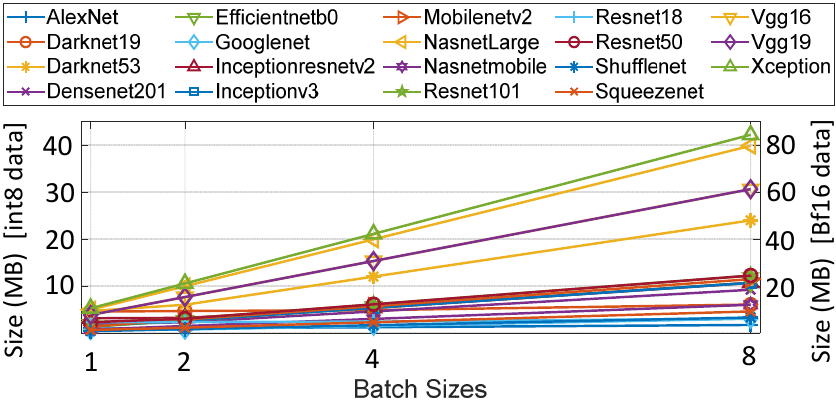}
	\caption {Required capacity of global buffer with varying batch sizes to avoid DRAM access during inference. }
	\label{fig:model_size_2}
\end{figure}

To fit a Conv. Layer data completely into the GLB, it needs the capacity to store the (i) input fmap (ifmap), (ii) filter weights, and (iii) output fmap (ofmap) of current layer. Fig. \ref{fig:model_size_2} shows the required GLB size for 19 widely-used deep learning models in int8 (left Y-axis) and BF16 (right Y-axis) for different batch sizes. For smaller batch-size (i.e., $\leq2$), a maximum of 12MB of GLB would be enough for int8 datatype. With 12MB on-chip GLB memory, most of the models, except a few (e.g., Darknet53, VGG19, Nasnetlarge, Xception, etc.), can support larger batch-sizes such as 8.  For BF16, 12MB would suffice for batch size 1 for all models. If pruned models \cite{aa1} are used, batch of more images can be fit into the GLB. For high-performance accelerators that operate with larger batches of data, the GLB size can be further increased.

\begin{figure}[h!]
	\centering
	\includegraphics[scale=0.97]{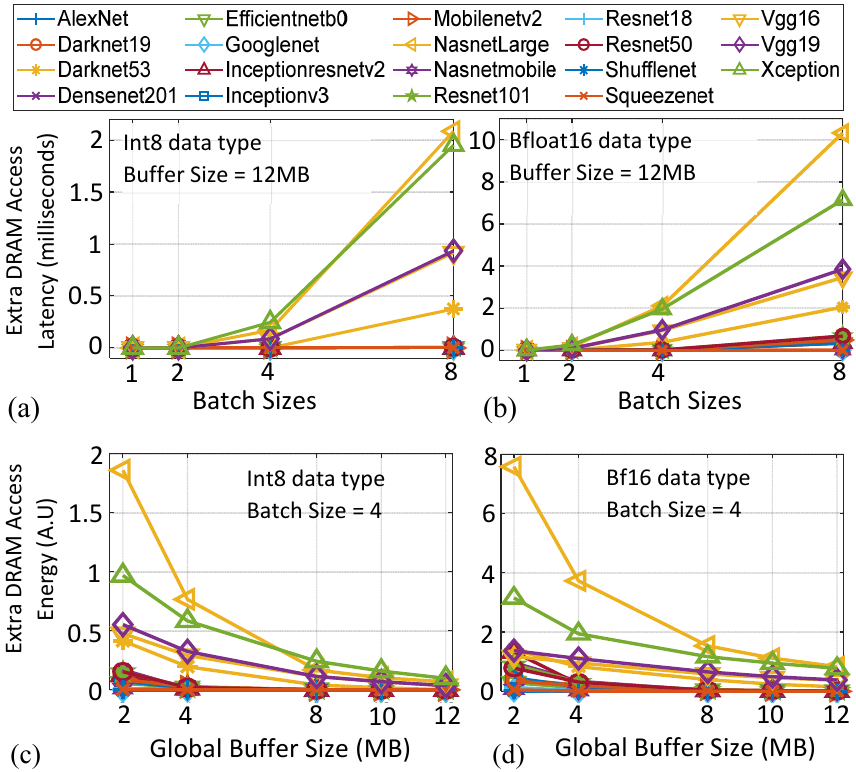}
	\caption {For Conv. layers, total extra DRAM access latency for varying batch sizes: (a) int8, (b) BF16 data types; total extra DRAM access energy for varying GLB size: (a) int8, (b) BF16 data types.}
	\label{fig:model_size_3}
\end{figure}

When a Conv. layer data - ifmap, weight, and ofmap - do not fit into GLB at one attempt, extra DRAM accesses are needed, incurring extra energy and latency. Fig. \ref{fig:model_size_3} (a) shows, if a GLB of 12MB is used, even larger batch sizes, such as 8, the extra DRAM access-related latency is zero for most of the models (int8 case), and around $2ms$ for few models. For BF16 datatype, the extra DRAM access latency increases slightly but is within $10ms$. Fig. \ref{fig:model_size_3} (c) depicts that if the GLB size is 12MB, for most of the models in int8 datatype extra DRAM access-associated energy reduces to zero. For BF16 datatype, most models would need a few extra DRAM accesses (Fig. \ref{fig:model_size_3} (d)). The DRAM access energy and latency were calculated for dual-channel DDR4-2933 DRAM with 64bit data bus.

\subsection{Memory Retention Time Estimation for AI Models and Accelerator Architecture}
\begin{figure}[h!]
	\centering
	\includegraphics[scale=0.97]{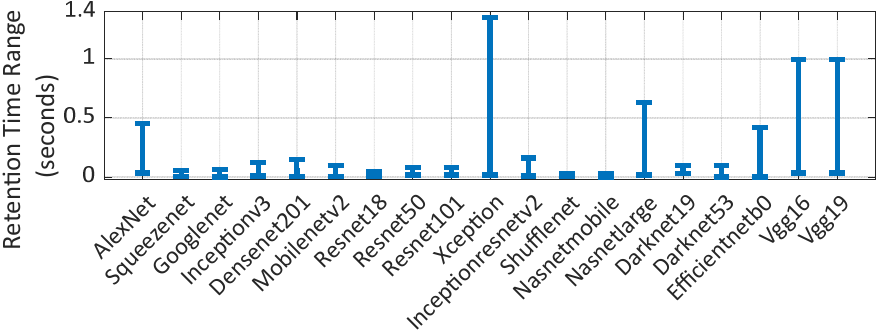}
	\caption {Global buffer retention time range for 42x42 MAC array (Bfloat16 hardware, CLK details in Table II) and batch size 16. }
	\label{fig:model_size_4}
\end{figure}

\begin{figure}[h!]
	\centering
	\includegraphics[scale=0.98]{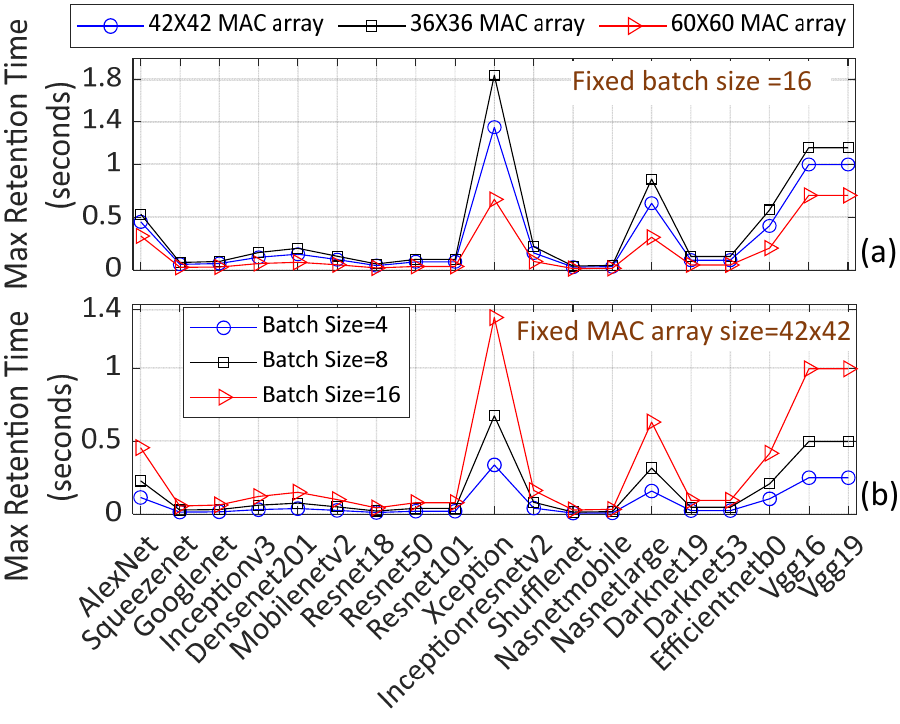}
	\caption {The required retention time of MRAM global buffer for Bfloat16 hardware (CLK cycles and frequency given in Table II), (a) varying MAC array capacity. (b) varying batch sizes. }
	\label{fig:model_size_5}
\end{figure}

The data retention time in GLB for the models (in BF16 datatype) are calculated using Equations \ref{eq:conv_Exp}-\ref{eq:Tret_conv_fc} (Section III) and the post-layout timing results from the implementation of our proposed reconfigurable accelerator core at 14nm technology (Table II). The results for 42\texttimes42 MAC array and batch size 16, presented in Fig. \ref{fig:model_size_4}, show that the maximum data retention time in GLB for all models is less than $1.5s$ where most models have retention time less than $0.5s$. The retention time goes even smaller (in $ms$ range) for int8 datatype as the clock cycle reduces significantly (usually 1-2 clock cycles) in int8 version hardware. Fig. \ref{fig:model_size_5} (a) shows the maximum retention time for all models (in BF16 datatype) for fixed batch size 16 and varying MAC array sizes, whereas Fig. \ref{fig:model_size_5} (b) shows the maximum retention time needed for a fixed MAC array size of 42\texttimes42 for varying batch sizes. From the figures, it is evident that further reduction in retention time can be achieved by using the proper combination of batch and MAC array sizes.

\begin{figure*}[h!]
	\centering
	\includegraphics[width=7.2in,height=1.3in]{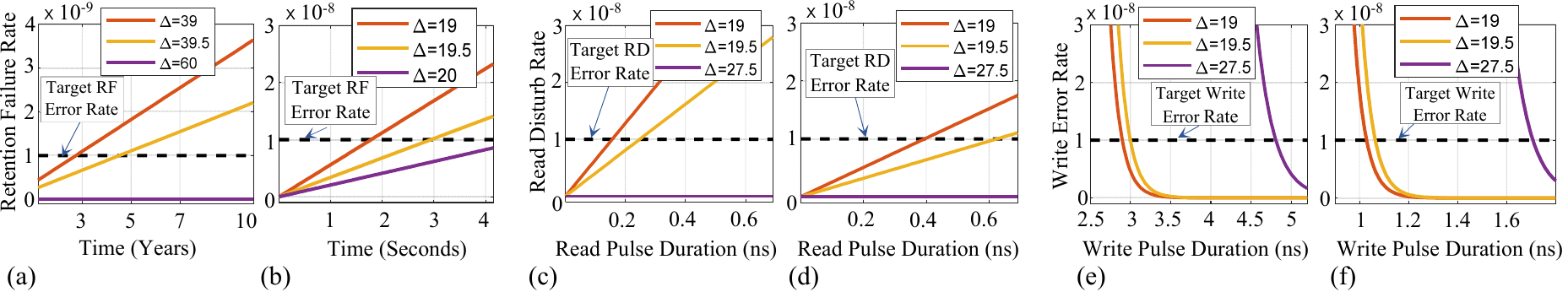}
	\caption {(a) Thermal stability ($\Delta$) scaling for 3 years retention time (for pre-trained weight storage NVM application).  (b) $\Delta$ and retention time scaling for accelerator's global buffer memory design. (c), (d) With scaled $\Delta$, read pulse width scaling while ensuring RD BER is within limit. (e), (f) Write latency scaling with $\Delta$, within target write error rate. Note: (c), (e) uses base-case (10yrs ret. time) from \cite{imec}, and (d), (f) from \cite{intel}. Target BER is chosen to ensure no accuracy impact on AI tasks \cite{ares}.}
	\label{fig:all_stt}
	\vspace{-0.1in}
\end{figure*}

\begin{figure}[h!]
	\centering
	\includegraphics[scale=1]{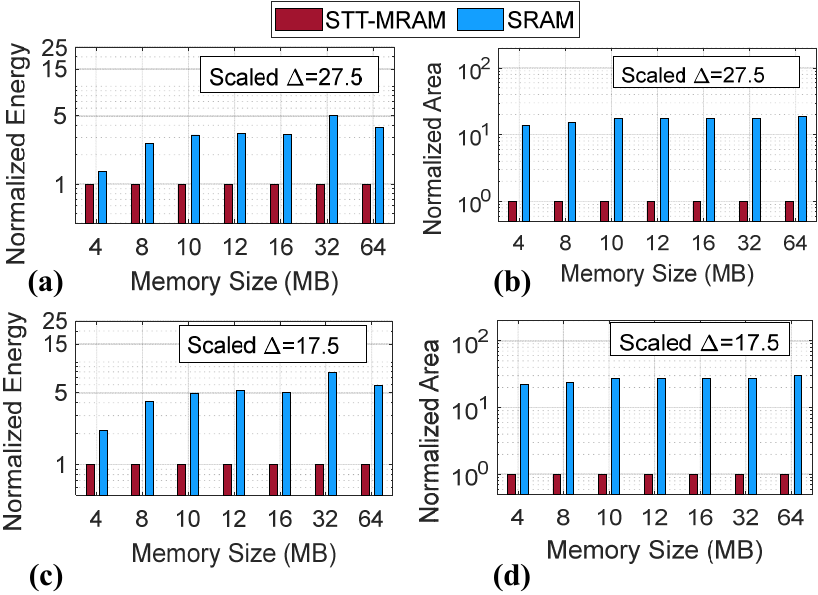}
	\caption {Energy and area comparison of SRAM and STT-MRAM for various sizes. $\Delta$ scaled: (a),(b) for global buffer; (c), (d) for eMRAM banks to store lower half (i.e., LSB groups) of weight/fmap bits.}
	\label{fig:area_energy}
\end{figure}

\subsection{Customizing STT-MRAM for AI Accelerator}
Using Equation (\ref{eq:ret}), we analyzed the impact of thermal stability factor ($\Delta$) on retention time within certain Bit Error Rates (BER). To identify the target BER of STT-RAM for applications in pre-trained weight storage and global buffer (GLB) memory, we first analyzed the size of modern AI models. From Fig. \ref{fig:model_size} (a), it can be seen that a few hundred MBs would be enough to store the pre-trained weights, within this memory capacity we choose BER in the order of $10^{-9}$ (i.e., 1 bit-flip per 1 billion bits). Given the worst-case cumulative BER can occur from Retention Failure (RF), Read Disturb (RD), and Write Error (WE), the worst-case bit-flips for VGG16 at this BER is about 12 bits. This BER is negligible and cannot make any impact on the AI task's accuracy \cite{ares}. Fig. \ref{fig:all_stt} (a) shows that with $\Delta=39$  we can ensure the loaded pre-trained weight will successfully remain in the accelerator for about 3 years at this target BER, which is enough given that AI models are replaced frequently with better ones. To address Process variation and runtime Temperature fluctuation, we chose $\sigma=2.1\%$ of mean,  $T_{hot}$=120$\degree$C ($393K$) and $T_{cold}$=-20$\degree$C ($253K$) in the Equations \ref{eq:delta_min} and \ref{eq:delta_max} as discussed in Section IV(C) and adjust $\Delta=39$ to $\Delta_{PT\_GB}=55$ after guard-banding.

For GLB memory, we can lower the $\Delta$ and retention time much lower according to the average occupancy time of weight and input/output fmaps in the accelerator's GLB memory. Also, since this memory size is within few tens of MB (e.g., 12MB), we can increase the BER  to $10^{-8}$, which will cause  less than 3 bit-flips in the worst-case (i.e., considering BER from Retention Failure (RF), Read Disturb (RD), and Write Error (WE)) at this memory size. The  accuracy impact of deep learning models at this BER and memory size is negligible \cite{ares}. In Fig. \ref{fig:all_stt} (b), at scaled $\Delta=19.5$ we can achieve 3 seconds of retention time (which is much higher than the minimum required as shown in Fig. \ref{fig:model_size_5}) at the target BER of $10^{-8}$.

Next, we analyze the impact of scaling $\Delta$ on the read pulse width. If the read pulse width is large then the chances of RD increase. Moreover, with scaling $\Delta=19.5$ (after Guardbanding $\Delta_{PT\_GB}=27.5$), the required read current also decreases. As a result, a significant reduction in read energy is also possible. In our study of  $\Delta$ scaling impact of read/write latency, as the base-case STT-MARM we used the chip-implemented (for 10 years retention) data of \cite{intel,imec}.  Fig. \ref {fig:all_stt} (c), (e) uses base-case (i.e., $\Delta=60$) from \cite{imec}, and (d), (f) from \cite{intel}.  With the scaling of retention time, the write latency only scales as a factor of $ln (\Delta)$, to further decrease the write latency we can use the write current as another knob as discussed in Section IV. The write latency scaling are shown in Fig. \ref {fig:all_stt}  (e), (f).

We used Destiny memory modeling tool \cite{destiny} to compare STT-MRAM area and energy with SRAM while $\Delta$ is scaled down. Although theoretically, STT-MRAM has a minimum area of $6F^2$, however, silicon results show that MRAM area scaled by 70\% compared to SRAM at 14nm node \cite{imec}. We modified the Destiny tool to incorporate this silicon observation. The results for scaled $\Delta$ at 14nm technology node are shown in Fig. \ref{fig:area_energy}. We see a significant advantage from STT-MRAM beyond 4MB capacity. Compared to SRAM, the area scales by more than ten times at iso-memory-capacity (Fig. \ref{fig:area_energy} (b),(d)). Similarly, for STT-MRAM the relative energy efficiency improves as the memory capacity increases (Fig. \ref{fig:area_energy} (a), (c)). These results imply STT-MRAM can offer significant performance gains at future high-performance AI accelerators that will use large on-chip buffer memory.

\subsection{Energy Optimization with Variable Retention MRAM Banks}
\begin{figure}[h]
	\centering
	\includegraphics[width=3.45in,height=1.23in]{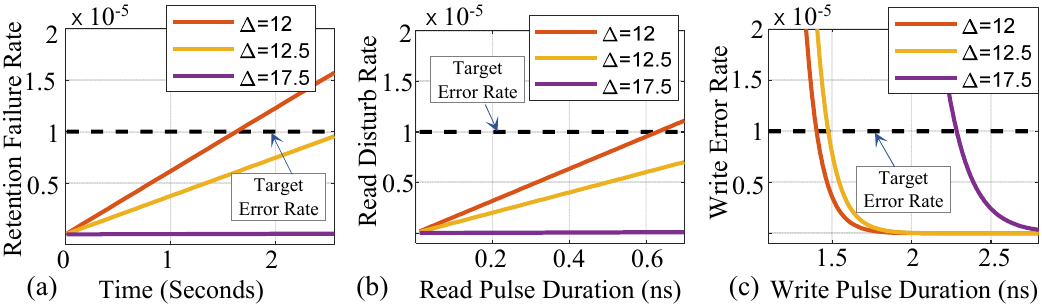}
	\vspace{-0.05in}
	\caption {$\Delta$ scaling with relaxed BER for LSB bit groups. (a) Retention, (b) Read, and (b) Write latency within target BER. (Base case, $\Delta=60$, data modeled after \cite{intel}).}
	\label{fig:approx_delta}
	\vspace{-0.0in}
\end{figure}

We further improved the efficiency in \textit{STT-AI Ultra} accelerator with two separate MRAM banks of, $\Delta=19.5$ (after PT guard-band $\Delta_{PT\_GB}=27.5$), and $\Delta=12.5$ ($\Delta_{PT\_GB}=17.5$). The first half of the weight/fmap bits are considered significant (MSB group) and stored in   $\Delta_{PT\_GB}=27.5$ bank, and the rest of the LSB groups in $\Delta_{PT\_GB}=17.5$ bank. For the LSB group at $\Delta_{PT\_GB}=17.5$  we relaxed the BER to $10^{-5}$ as shown in Fig. \ref{fig:approx_delta}. The relative gains in energy and area at  $\Delta_{PT\_GB}=17.5$ are shown in Fig. \ref{fig:area_energy} (c), (d).

\subsection{Optimizing Energy with Scratchpad for Partial Ofmaps}
\begin{figure}[h!]
	\centering
	\includegraphics[scale=1.05]{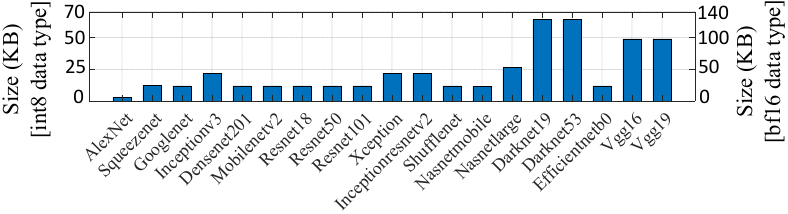}
	\caption {Maximum size of partial ofmaps. }
	\label{fig:model_size_ofmap}
\end{figure}

\begin{figure}[h!]
	\centering
	\includegraphics[scale=0.92]{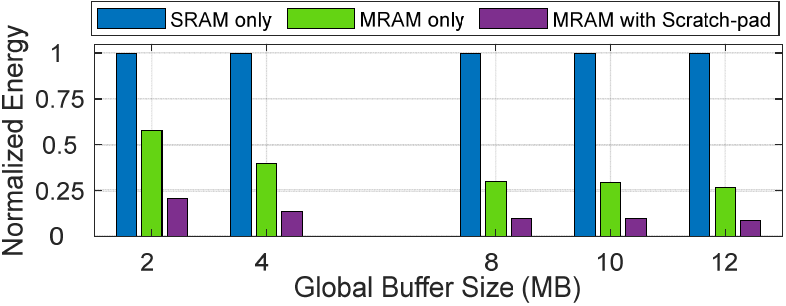}
	\caption {Comparison of buffer memory energy dissipation for, (i) SRAM, (ii) MRAM, and (iii) MRAM with scratch pad architectures. }
	\label{fig:stt_dataflow}
\end{figure}

Our simulation results show that for STT-MRAM the write energy is about 70\% more than the read energy at scaled $\Delta$. As described in Section IV (D), using a small SRAM scratchpad for writing the intermediate partial ofmaps instead of the MRAM can significantly reduce the write frequency and save energy. Fig. \ref{fig:model_size_ofmap} shows the partial ofmap size distribution. For BF16 data type, we see that 52KB (26KB for int8) scratchpad will fit most of the models in one attempt. The normalized energy improvements of proposed scratchpad-assisted MRAM system is shown in Fig. \ref{fig:stt_dataflow} for ResNet-50 model and 14nm technology. 

\subsection{Accelerator Implementation}
\begin{figure}[h]
	\centering
		\includegraphics[scale=.94]{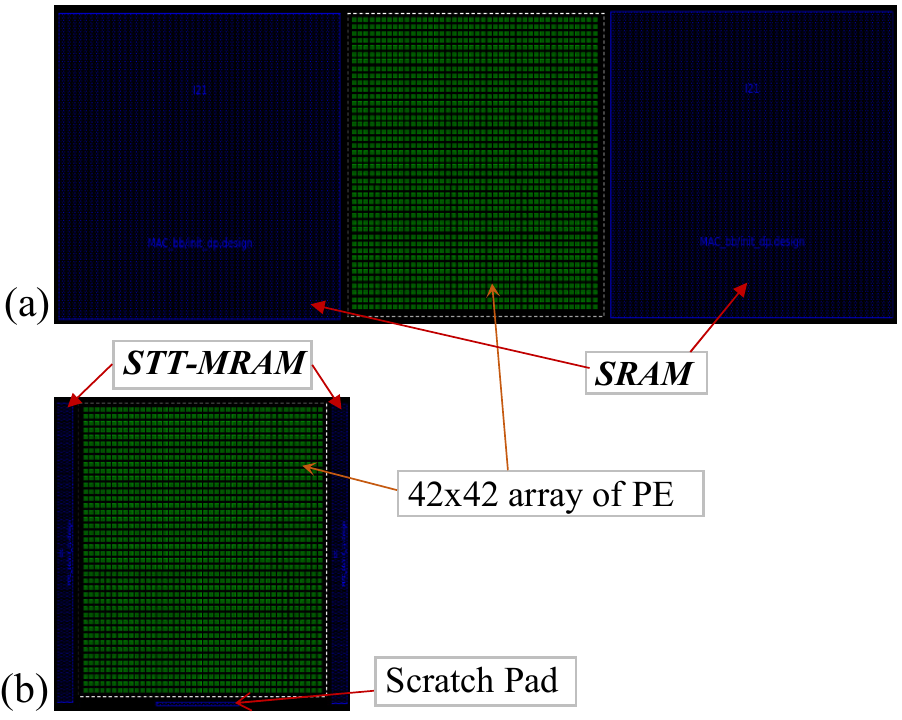}
	\vspace{-0.05in}
	\caption {Top level floorplan view from ICC2. Accelerator designed with, (a) 12MB SRAM. (b) 12MB STT-MRAM with scratchpad.}
	\label{fig:floorplan}
	\vspace{-0.0in}
\end{figure}

We implemented our AI accelerator architecture with reconfigurable cores (i.e., in Fig. 3),  at RTL level using BF16 hardware as BF16 can support both inference and training. We used Synopsys 14nm standard cell library \cite{saed14} to complete synthesis, and place and route of the design. The post-layout CLK cycle data for the PE/MAC are shown in Table II. The top-level view of physical design from ICC2 tool \cite{saed14} is shown in Fig. \ref{fig:floorplan}.  We used Synopsys 14nm memory compiler to create the SRAMs for our baseline accelerator. Results from post-layout and timing-closed accelerator design are shown in Table III, where Row 7 shows the area and power for the base-line accelerator with 12MB global buffer memory. Next, to implement MRAM based \textit{STT-AI} accelerator, we estimated areas and power data from the Destiny \cite{destiny} tool at scaled $\Delta$ and modeled those as blackbox in the physical design part in Synopsys ICC2 \cite{saed14} for 14nm node. The 52KB SRAM Scratchpad is divided into two banks with individual CLK/power gating. Rows 4 and 8 show that the \textit{STT-AI} accelerator offers significant area and leakage energy savings. The \textit{STT-AI Ultra} accelerator achieves further improvements in power and area as shown in Row 9 in Table III.

\begin{table}[]
\centering
\caption {Reconfigurable PE core details (Bfloat16 hardware, and synthesized with 14nm standard cell library \cite{saed14})}
\renewcommand{\arraystretch}{1.25}
\begin{tabular}{|c|c|c|}
\hline
\textit{Reconfigurable Core Mode} & \textit{CLK Freq} & \textit{Required CLK Cycles} \\ \hline
Systolic Core (1 MAC)                        & 1 GHz             & 11                           \\ \hline
Conv. Core (3 MAC)                           & 1 GHz             & 17                           \\ \hline
\end{tabular}
\end{table}

\begin{table}[]
\centering
\caption {Accelerator Design Details at 14nm }
\vspace{-0.05in}
\scriptsize
\renewcommand{\arraystretch}{1.25}
\begin{tabular}{|c|c|c|c|c|}
\hline
Module                                                                          & Details                                                                                                                  & \begin{tabular}[c]{@{}c@{}}Area\\  ($mm^2$)\end{tabular} & \begin{tabular}[c]{@{}c@{}}Dynamic \\ Power\\  ($mW$)\end{tabular} & \begin{tabular}[c]{@{}c@{}}Leakage\\ Power \\ ($mW$)\end{tabular} \\ \hline
\begin{tabular}[c]{@{}c@{}}Functional\\ Core\end{tabular}                       & \begin{tabular}[c]{@{}c@{}}Reconfigurable core\\  with 42x42 MACs\end{tabular} & 4.08                                                  & 954                                                            & 0.91                                                           \\ \hline
\begin{tabular}[c]{@{}c@{}}SRAM \\ Block\end{tabular}                           & \begin{tabular}[c]{@{}c@{}}12 MB SRAM global\\ memory\end{tabular}                                                        & 16.2                                                   & 48.98                                                            & 0.21                                                              \\ \hline
\begin{tabular}[c]{@{}c@{}}STT-MARM \\ ($\Delta$=27.5)\end{tabular}                      & \begin{tabular}[c]{@{}c@{}}12 MB MRAM\\ global memory\end{tabular}                                                    & 1.01                                                  & 17.61                                                            & 0.08                                                              \\ \hline
\begin{tabular}[c]{@{}c@{}}STT-MRAM\\  ($\Delta$=17.5, $\Delta$=27.5)\end{tabular}                & \begin{tabular}[c]{@{}c@{}}6 MB MRAM ($\Delta$=17.5)\\ 6 MB MRAM ($\Delta$=27.5)\end{tabular}                                      & 0.93                                                  & 13.75                                                             & 0.06                                                              \\ \hline
\begin{tabular}[c]{@{}c@{}}SRAM ScratchPad\\ (for MRAM)\end{tabular}                & \begin{tabular}[c]{@{}c@{}}52 KB (two 26KB blocks \\ with CLK/power gating)\end{tabular}                                      & 0.069                                                  & 0.2                                                             & 8E-4                                                              \\ \hline
\begin{tabular}[c]{@{}c@{}}\textit{Baseline} \\ \textit{Accelerator} \\ (SRAM-based)\end{tabular} & \begin{tabular}[c]{@{}c@{}}Functional Core and\\  SRAM  (Row 3 above)\end{tabular}                                       & 20.28                                                  & 1003                                                            & 1.13                                                             \\ \hline
\begin{tabular}[c]{@{}c@{}}\textit{STT-AI} \\ \textit{Accelerator}\end{tabular}                   & \begin{tabular}[c]{@{}c@{}}Functional core and\\  STT-RAM (Row 4 above)\end{tabular}                                     & 5.09                                                  & 972                                                           & 0.99                                                              \\ \hline
\begin{tabular}[c]{@{}c@{}}\textit{STT-AI Ultra} \\ \textit{Accelerator  }\end{tabular}            & \begin{tabular}[c]{@{}c@{}}Functional Core and \\ STT-RAM (Row 5 above)\end{tabular}                                     & 5.0                                                  & 968                                                            & 0.98                                                              \\ \hline
\end{tabular}
\end{table}

\begin{figure}[h!]
	\centering
	\includegraphics[width=3.45in,height=1.05in]{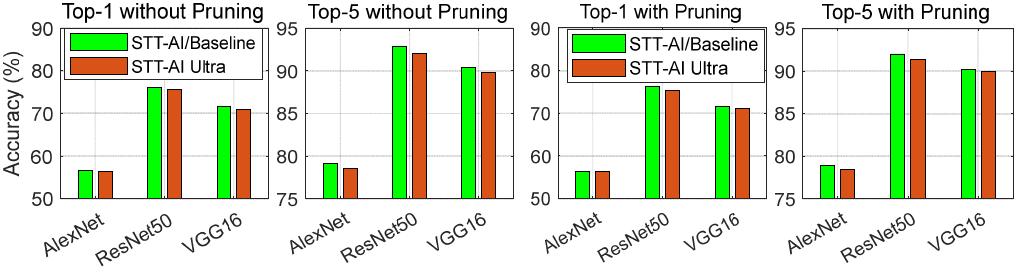}
	\caption { Top-1 and Top-5 accuracy comparisons for \textit{STT-AI/Baseline} and \textit{STT-AI Ultra} cases. No accuracy change for \textit{STT-AI/Baseline} cases, and negligible (less than 1\% normalized) accuracy change occurs on \textit{STT-AI Ultra} acclerator. Both original and pruned (at 50\% pruing rate) \cite{aa1} model results are shown.}
	\label{fig:approx_pyt}
\end{figure}

\subsection{Accelerator Performance with ImageNet Dataset}
Next, we modeled our hardware and STT-MRAM BERs in PyTorch \cite{pyt} and ran inference for pre-trained AlexNet, VGG16, and  ResNet-50 models with ImageNet benchmark to obtain Top-1 and Top-5 accuracy results. As expected, with STT-MRAM having $\Delta_{PT\_GB}=27.5$, there were no accuracy loss compared to the baseline SRAM version. However, for \textit{STT-AI Ultra} accelerator, with $BER=10^{-5}$ in half of the bits (LSB group in lower $\Delta$ STT-MRAM bank), we observed negligible (less than 1\% normalized) accuracy loss as shown in Fig. \ref{fig:approx_pyt}.

\section{Related Work}
\vspace{-0.02in}
\label{sec:Related}
Over the last decade, STT-MRAM technology has been extensively researched for its high-endurance, radiation hardness, non-volatility, and high-density memory properties.  The prior works on STT-MRAM applications can be broadly categorized into two domains: (i) Its use as the last-level cache memory in processors; (ii) Its application in emerging Process-in-Memory (PIM) based computing paradigm.

Several studies \cite{jain2017computing, yan2019icelia, anwar2020xbaropt, shi2020performance} have demonstrated the excellence of STT-MRAM over other NVM technologies in PIM setting due to complex and tunable resistance dynamics achieved through its spin-transfer torque mechanism and simultaneous access to multiple word-lines of the same array. \cite{yan2019icelia} proposed an STT-MRAM based crossbar arrays where the internal resistance states - which were used to mimic the weights of the models - of STT-MRAM were tuned to support non-uniform quantization. This study provided energy efficiency and loss reduction, however, additional circuitry, Digital to Analog Converter (DAC) and Analog to Digital Converter (ADC) were needed to support the PIM workflow. Some studies, such as, \cite{mrima} used the PIM architecture, where the MAC operation was simplified to addition/subtraction, and bit-wise operation by manipulating the models' parameters. \cite{shi2020performance} mapped the LeNet5 model to a synaptic cross-bar array of STT-MRAM memory cells for inference and showed improved performance in terms of area, leakage power, energy over SRAM for 65nm to 7nm technology node. However, major challenges of PIM over conventional Deep Learning/AI are - (i) requirements of additional hardware circuitry, such as DAC, ADC, which results in area overhead; (2) quantization of weights to be represented with fewer bits resulting in lower precision; (3) in digital PIM, extra manipulation of models' algorithm is needed to replace MAC with the bit-wise operation; and (4) in most of the cases, PIM is only suitable for inference-only applications. Although PIM-based analog architectures provide fast execution, the performance, energy efficiency, and reliability of analog PIM still lags behind the state-of-the-art DNN/AI models and their corresponding hardware accelerators \cite{aa1,tpu,dac_19}.

While some research leveraged the scalable property of thermal stability factor of STT-MRAM to replace SRAM-based cache memory, others used the error tolerance property of certain applications and designed STT-MRAM based energy-efficient cache with approximate storage. In \cite{uva}, the retention time of STT-RAM was scaled to implement cache memory that can compete with SRAM-based caches, and DRAM-like refresh was used to compromise the ultra-low retention time. In \cite{approx_mram, sayed2020approximate} STT-MRAM based approximate cache was proposed to exploit the error-resilience property of some specific applications.  Unlike previous studies, \cite{tahoori_1} proposed a hybrid STT-MRAM design for cache for different applications depending on the run-time requirements without compromising any reliability degradation. In \cite{scale_thermal}, area-and-retention-time-scaled STT-MRAM was presented as DRAM replacement.

In \cite{gtech}  a hybrid of SRAM and 3D-stacked STT-MRAM based AI accelerator was proposed for real-time learning where eMRAM acted as weight storage memory for infrequently accessed and updated layers, such as all convolutional layers and first few fully connected layers for a Transfer Learning followed by Reinforcement Learning algorithm. However, due to the use of typical slow and write-power-hungry STT-MRAM, this study could not completely exploit STT-MRAM to substitute SRAM and eventually used SRAM for storing weights of the last few fully connected layers which are accessed and updated frequently in transfer learning-based reinforcement learning setting. 

In summary, prior notable research on STT-MRAM applications focused on implementing last-level cache memory, designing in-memory computing architectures, and high-capacity 3D-stacked memory as DRAM replacement for DNN hardware. Our work is the first to present a detailed analysis on the feasibility of using STT-MRAM as high bandwidth on-chip buffer memory in DNN/AI accelerator hardware that can offer much larger capacity at lower energy and area costs compared to SRAM. Moreover, for complete DNN model storage in edge inference devices, non-volatility relaxed STT-MRAM design is presented as an alternative to eflash which suffers from scaling limitations at advanced technology nodes.

\section{Conclusions}
\vspace{-0.02in}
\label{sec:conc}
In this paper, we demonstrated the design of highly efficient AI/Deep Learning accelerators that utilize emerging STT-MRAMs. Based on detailed design space exploration we designed the STT-MRAM based global buffer to minimize DRAM access latency and energy, as well as reduce the area and power of the MRAM buffer. We presented an innovative runtime-reconfigurable core optimized for both dot products and matrix multiplication in convolution and fully-connected layers, respectively. A scratchpad-assisted STT-MRAM global buffer design has been demonstrated that reduces the frequency of energy-dominant write operations of the partial ofmaps during convolution. Using actual data occupancy times in memory for AI tasks, we guide the STT-MRAM thermal stability factor scaling. We showed that with \textit{STT-AI} accelerator 75\% area and 3\% power savings are possible at iso-accuracy. Furthermore, with \textit{STT-AI Ultra}, 75.4\%, and 3.5\% savings in area and power, respectively, over regular SRAM-based accelerators at minimal accuracy trade-off.

\bibliographystyle{IEEEtran}
\tiny
\bibliography{reference}

\begin{thebibliography}{10}
\providecommand{\url}[1]{#1}
\csname url@samestyle\endcsname
\providecommand{\newblock}{\relax}
\providecommand{\bibinfo}[2]{#2}
\providecommand{\BIBentrySTDinterwordspacing}{\spaceskip=0pt\relax}
\providecommand{\BIBentryALTinterwordstretchfactor}{4}
\providecommand{\BIBentryALTinterwordspacing}{\spaceskip=\fontdimen2\font plus
\BIBentryALTinterwordstretchfactor\fontdimen3\font minus
  \fontdimen4\font\relax}
\providecommand{\BIBforeignlanguage}[2]{{%
\expandafter\ifx\csname l@#1\endcsname\relax
\typeout{** WARNING: IEEEtran.bst: No hyphenation pattern has been}%
\typeout{** loaded for the language `#1'. Using the pattern for}%
\typeout{** the default language instead.}%
\else
\language=\csname l@#1\endcsname
\fi
#2}}
\providecommand{\BIBdecl}{\relax}
\BIBdecl

\bibitem{market}
{G. Batra et al.}, ``{Artificial-intelligence hardware: New opportunities for
  semiconductor companies}.''\hskip 1em plus 0.5em minus 0.4em\relax McKinsey
  \& Company, 2019.

\bibitem{aa1}
V.~{Sze}, Y.~{Chen}, T.~{Yang}, and J.~S. {Emer}, ``{Efficient Processing of
  Deep Neural Networks: A Tutorial and Survey},'' in \emph{{Proc. of the
  IEEE}}, 2017.

\bibitem{tpu}
{N. Jouppi et al.}, ``{A Domain-Specific Architecture for Deep Neural
  Networks}.''\hskip 1em plus 0.5em minus 0.4em\relax ACM Communications, 2018.

\bibitem{eyeriss}
Y.~{Chen}, T.~{Krishna}, J.~S. {Emer}, and V.~{Sze}, ``{Eyeriss: An
  Energy-Efficient Reconfigurable Accelerator for Deep Convolutional Neural
  Networks},'' in \emph{IEEE JSSC}, 2017.

\bibitem{DNPU}
D.~{Shin}, J.~{Lee}, J.~{Lee}, J.~{Lee}, and H.~{Yoo}, ``{DNPU: An
  Energy-Efficient Deep-Learning Processor with Heterogeneous Multi-Core
  Architecture},'' in \emph{IEEE Micro}, 2018.

\bibitem{imec}
{S. Sakhare et al.}, ``{$J_{SW}$ of 5.5 $MA/cm^2$ and RA of 5.2-$\Omega$.$\mu
  m^2$ STT-MRAM Technology for LLC Application},'' in \emph{IEEE Transactions
  on Electron Devices}, vol.~67, no.~9, 2020, pp. 3618--3625.

\bibitem{memory_trend}
\BIBentryALTinterwordspacing
{H.-S. P. Wong et al.}, ``{Stanford Memory Trends},'' 2020. [Online].
  Available: \url{https://nano.stanford.edu/stanford-memory-trends}
\BIBentrySTDinterwordspacing

\bibitem{cb}
{S. Moore}, ``{Cerebras's Giant Chip Will Smash Deep Learning’s Speed
  Barrier}.''\hskip 1em plus 0.5em minus 0.4em\relax IEEE Spectrum, 2020.

\bibitem{ser}
{G. Li et al.}, ``{Understanding Error Propagation in Deep Learning Neural
  Network (DNN) Accelerators and Applications},'' in \emph{ACM Conference for
  High Performance Computing, Networking, Storage and Analysis}, 2017.

\bibitem{samsung}
A.~{Antonyan}, S.~{Pyo}, H.~{Jung}, and T.~{Song}, ``{Embedded MRAM Macro for
  eFlash Replacement},'' in \emph{IEEE ISCAS}, 2018.

\bibitem{dac_19}
H.~{Li}, M.~{Bhargav}, P.~N. {Whatmough}, and H.~{Philip Wong}, ``{On-Chip
  Memory Technology Design Space Explorations for Mobile Deep Neural Network
  Accelerators},'' in \emph{Design Automation Conference (DAC)}, 2019.

\bibitem{stt_ram_tsmc1}
{Q. Dong et al.}, ``{A 1Mb 28nm STT-MRAM with 2.8ns read access time at 1.2V
  VDD using single-cap offset-cancelled sense amplifier and in-situ
  self-write-termination},'' in \emph{{IEEE ISSCC}}, 2018.

\bibitem{intel}
{L. Wei et al.}, ``{A 7Mb STT-MRAM in 22FFL FinFET Technology with 4ns Read
  Sensing Time at 0.9V Using Write-Verify-Write Scheme and Offset-Cancellation
  Sensing Technique},'' in \emph{{IEEE ISSCC}}, 2019.

\bibitem{stt_ram_tsmc2}
{Y. Chih et al.}, ``{A 22nm 32Mb Embedded STT-MRAM with 10ns Read Speed, 1M
  Cycle Write Endurance, 10 Years Retention at 150°C and High Immunity to
  Magnetic Field Interference},'' in \emph{{IEEE ISSCC}}, 2020.

\bibitem{samsung_19}
{J. {Park} et al.}, ``A novel integration of stt-mram for on-chip hybrid memory
  by utilizing non-volatility modulation,'' in \emph{IEEE International
  Electron Devices Meeting (IEDM)}, 2019.

\bibitem{ibm_2019}
{G. {Hu} et al.}, ``Spin-transfer torque mram with reliable 2 ns writing for
  last level cache applications,'' in \emph{IEEE International Electron Devices
  Meeting (IEDM)}, 2019.

\bibitem{destiny}
M.~{Poremba}, S.~{Mittal}, D.~{Li}, J.~S. {Vetter}, and Y.~{Xie}, ``Destiny: A
  tool for modeling emerging 3d nvm and edram caches,'' in \emph{Design,
  Automation Test in Europe Conference Exhibition}, 2015.

\bibitem{stt_model}
A.~{Raychowdhury}, D.~{Somasekhar}, T.~{Karnik}, and V.~{De}, ``{Design space
  and scalability exploration of 1T-1STT MTJ memory arrays in the presence of
  variability and disturbances},'' in \emph{IEEE IEDM}, 2009.

\bibitem{bfloat}
S.~{Wang} and P.~{Kanwar}, ``Bfloat16: The secret to high performance on cloud
  tpus,'' in \emph{Google Cloud Blog}, 2019.

\bibitem{bf16_fb}
{D. Kalamkar et al.}, ``A study of bfloat16 for deep learning training,'' in
  \emph{arXiv:1905.12322}, 2019.

\bibitem{stt_eqn1}
{A V Khvalkovskiy et al.}, ``{Basic principles of {STT}-{MRAM} cell operation
  in memory arrays},'' in \emph{Journal of Physics D: Applied Physics}, 2013.

\bibitem{stt_eqn2}
{Z Diao et al.}, ``{Spin-transfer torque switching in magnetic tunnel junctions
  and spin-transfer torque random access memory},'' in \emph{Journal of
  Physics: Condensed Matter}.\hskip 1em plus 0.5em minus 0.4em\relax {IOP}
  Publishing, 2007.

\bibitem{scale_thermal}
Y.~Jin, M.~Shihab, and M.~Jung, ``{Area, Power, and Latency Considerations of
  STT-MRAM to Substitute for Main Memory},'' 2014.

\bibitem{tdk_18}
{J. {Iwata-Harms} et al.}, ``High-temperature thermal stability driven by
  magnetization dilution in cofeb free layers for spin-transfer-torque magnetic
  random access memory,'' in \emph{Nature Scientific Reports}, 2018.

\bibitem{ares}
B.~{Reagen}, U.~{Gupta}, L.~{Pentecost}, P.~{Whatmough}, S.~K. {Lee},
  N.~{Mulholland}, D.~{Brooks}, and G.~{Wei}, ``Ares: A framework for
  quantifying the resilience of deep neural networks,'' in \emph{Design
  Automation Conference (DAC)}, 2018.

\bibitem{saed14}
\BIBentryALTinterwordspacing
``{Synopsys Inc.}'' [Online]. Available: \url{https://www.synopsys.com//}
\BIBentrySTDinterwordspacing

\bibitem{pyt}
\BIBentryALTinterwordspacing
``{PyTorch}.'' [Online]. Available: \url{https://pytorch.org/}
\BIBentrySTDinterwordspacing

\bibitem{jain2017computing}
S.~Jain, A.~Ranjan, K.~Roy, and A.~Raghunathan, ``Computing in memory with
  spin-transfer torque magnetic ram,'' in \emph{IEEE Transactions on Very Large
  Scale Integration (VLSI) Systems}, 2017.

\bibitem{yan2019icelia}
H.~Yan, H.~R. Cherian, E.~C. Ahn, X.~Qian, and L.~Duan, ``icelia: a full-stack
  framework for stt-mram-based deep learning acceleration,'' \emph{IEEE
  Transactions on Parallel and Distributed Systems}, 2019.

\bibitem{anwar2020xbaropt}
A.~Anwar, A.~Raychowdhury, R.~Hatcher, and T.~Rakshit, ``Xbaropt-enabling
  ultra-pipelined, novel stt mram based processing-in-memory dnn accelerator,''
  in \emph{IEEE International Conference on Artificial Intelligence Circuits
  and Systems (AICAS)}, 2020.

\bibitem{shi2020performance}
Y.~Shi, S.~Oh, Z.~Huang, X.~Lu, S.~H. Kang, and D.~Kuzum, ``Performance
  prospects of deeply scaled spin-transfer torque magnetic random-access memory
  for in-memory computing,'' in \emph{IEEE Electron Device Letters}, 2020.

\bibitem{mrima}
S.~{Angizi}, Z.~{He}, A.~{Awad}, and D.~{Fan}, ``Mrima: An mram-based in-memory
  accelerator,'' in \emph{IEEE Transactions on Computer-Aided Design of
  Integrated Circuits and Systems}, 2020.

\bibitem{uva}
C.~W.~S. et~al., ``{Relaxing non-volatility for fast and energy-efficient
  STT-RAM caches},'' in \emph{IEEE 17th International Symposium on High
  Performance Computer Architecture}, 2011.

\bibitem{approx_mram}
{A. {Ranjan} and S. {Venkataramani} and Z. {Pajouhi} and R. {Venkatesan} and K.
  {Roy} and A. {Raghunathan}}, ``{STAxCache: An approximate, energy efficient
  STT-MRAM cache},'' in \emph{DATE}, 2017.

\bibitem{sayed2020approximate}
N.~Sayed, R.~Bishnoi, and M.~B. Tahoori, ``Approximate spintronic memories,''
  in \emph{ACM Journal on Emerging Technologies in Computing Systems (JETC)},
  2020.

\bibitem{tahoori_1}
N.~Sayed, L.~Mao, R.~Bishnoi, and M.~B. Tahoori, ``Compiler-assisted and
  profiling-based analysis for fast and efficient stt-mram on-chip cache
  design,'' in \emph{ACM Transactions on Design Automation of Electronic
  Systems (TODAES)}, 2019.

\bibitem{gtech}
I.~{Yoon}, M.~A. {Anwar}, R.~V. {Joshi}, T.~{Rakshit}, and A.~{Raychowdhury},
  ``{Hierarchical Memory System With STT-MRAM and SRAM to Support Transfer and
  Real-Time Reinforcement Learning in Autonomous Drones},'' in \emph{IEEE
  Journal on Emerging Topics}, 2019.

\end{thebibliography}

\begin{IEEEbiography}[{\includegraphics[width=1in,height=1.25in,clip,keepaspectratio]{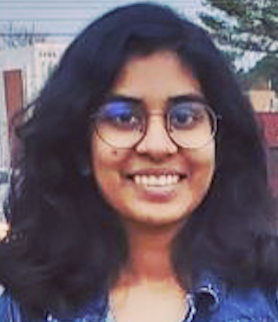}}]{Kaniz Mishty} received the B.S. degree in Electronics and Communication Engineering from Khulna University of Engineering and Technology, Bangladesh, in 2018. She is currently working towards her Ph.D. degree in ECE at Auburn University, AL, USA. Her current research interests are energy and area efficient VLSI system design, AI/Neuromorphic hardware design and AI/ML in CAD. As a summer intern, she will work on incorporating AI/Machine Learning in ASIC design flows at Qualcomm, Santa Clara.
\end{IEEEbiography}

\begin{IEEEbiography}[{\includegraphics[width=1in,height=1.25in,clip,keepaspectratio]{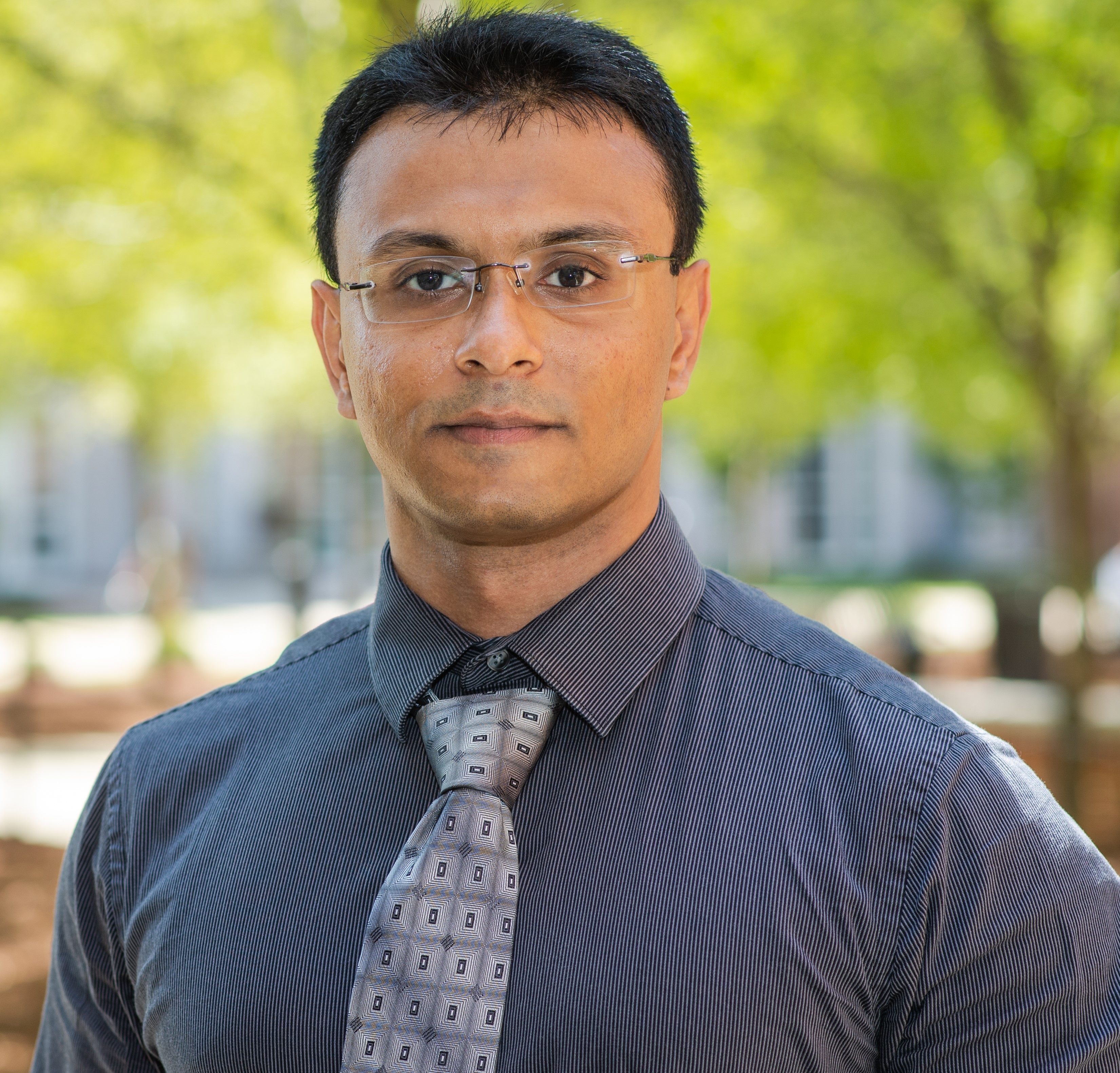}}]{Mehdi Sadi} (S'12-M'17) is currently an Assistant Professor at the Department of Electrical and Computer Engineering (ECE) at Auburn University, Auburn, AL.  Dr. Sadi  earned his PhD in ECE from  University of Florida, Gainesville, USA in 2017, MS from University of California at Riverside, USA in 2011 and BS from Bangladesh University of Engineering and Technology in 2010.   Prior to joining Auburn University, he was a Senior R\&D SoC Design Engineer in the Xeon server CPU design team at Intel Corporation in Oregon. Dr. Sadi`s research focus is on developing algorithms and Computer-Aided-Design (CAD) techniques for implementation, design, reliability, and security of AI, and brain-inspired computing hardware. His research also spans into developing Machine Learning/AI enabled design flows for  System-on-Chip (SoC), and Design-for-Robustness for safety-critical AI hardware systems. He has published more than 20 peer-reviewed research papers. He was the recipient of Semiconductor Research Corporation best in session award and Intel Xeon Design Group recognition awards.
\end{IEEEbiography}

\end{document}